\documentclass[aps,superscriptaddress,amsmath,amssymb,twocolumn,showpacs,floatfix,reprint]{revtex4-2}
\usepackage{bm}
\usepackage{graphicx}
\usepackage[colorlinks=true, urlcolor=blue, linkcolor=blue, citecolor=blue, pdftex]{hyperref}
\usepackage{float}
\usepackage[usenames,dvipsnames]{xcolor}
\usepackage{comment}
\usepackage{braket}
\usepackage{booktabs}
\usepackage[nice]{nicefrac}

\definecolor{C0}{HTML}{1f77b4}
\definecolor{C1}{HTML}{ff7f0e}
\definecolor{C2}{HTML}{2ca02c}
\definecolor{C3}{HTML}{d62728}
\definecolor{C4}{HTML}{9467bd}
\definecolor{C5}{HTML}{8c564b}

\usepackage{lineno}
\usepackage{todonotes}

\begin{document}

\title{Transformer Wave Function for two dimensional frustrated magnets: \\
emergence of a Spin-Liquid Phase in the Shastry-Sutherland Model}

\author{Luciano Loris Viteritti}
\thanks{These authors contributed equally.}
\affiliation{Dipartimento di Fisica, Universit\`a di Trieste, Strada Costiera 11, I-34151 Trieste, Italy}
\author{Riccardo Rende}
\thanks{These authors contributed equally.}
\affiliation{International School for Advanced Studies (SISSA), Via Bonomea 265, I-34136 Trieste, Italy}
\author{Alberto Parola}
\affiliation{Dipartimento di Scienza e Alta Tecnologia, Universit\`a dell'Insubria, Via Valleggio 11, I-22100 Como, Italy}
\author{Sebastian Goldt}
\affiliation{International School for Advanced Studies (SISSA), Via Bonomea 265, I-34136 Trieste, Italy}
\author{Federico Becca}
\affiliation{Dipartimento di Fisica, Universit\`a di Trieste, Strada Costiera 11, I-34151 Trieste, Italy}
\thanks{Correspondence should be addressed to lucianoloris.viteritti@phd.units.it and rrende@sissa.it}

\date{\today}

\begin{abstract}
Understanding quantum magnetism in two-dimensional systems represents a lively branch in modern condensed-matter physics. In the presence of competing super-exchange 
couplings, magnetic order is frustrated and can be suppressed down to zero temperature. Still, capturing the correct nature of the exact ground state is a highly 
complicated task, since energy gaps in the spectrum may be very small and states with different physical properties may have competing energies. Here, we introduce 
a variational {\it Ansatz} for two-dimensional frustrated magnets by leveraging the power of representation learning. The key idea is to use a particular deep neural 
network with real-valued parameters, a so-called Transformer, to map physical spin configurations into a high-dimensional feature space. Within this abstract space, 
the determination of the ground-state properties is simplified and requires only a shallow output layer with complex-valued parameters. We illustrate the efficacy
of this variational {\it Ansatz} by studying the ground-state phase diagram of the Shastry-Sutherland model, which captures the low-temperature behavior of
SrCu$_2$(BO$_3$)$_2$ with its intriguing properties. With highly accurate numerical simulations, we provide strong evidence for the stabilization of a spin-liquid 
between the plaquette and antiferromagnetic phases. In addition, a direct calculation of the triplet excitation at the $\Gamma$ point provides compelling evidence 
for a gapless spin liquid. Our findings underscore the potential of Neural-Network Quantum States as a valuable tool for probing uncharted phases of matter, and 
open up new possibilities for establishing the properties of many-body systems.
\end{abstract}

\maketitle

\section{Introduction}

Since the discovery of the fractional quantum Hall effect~\cite{tsui1982} and its description by the Laughlin wave function~\cite{laughlin1983}, a growing interest 
has developed around unconventional phases of matter, i.e., the ones that escape perturbative or mean-field approaches. In this sense, the hunt for spin liquids is 
of fundamental importance in Mott insulators, where localized spins determine the low-temperature properties. On geometrically frustrated lattices, it is not possible 
to minimize simultaneously all the interactions among the spins and, therefore, magnetic order could be suppressed, even at zero temperature. In this case, spins 
are highly entangled and the resulting ground-state wave function shows unconventional properties~\cite{savary2017}. However, most of the theoretical models that 
have been proposed to support quantum spin liquids are still unresolved, and their phase diagrams are not well established except for specific points (that usually 
give trivial states). One notable exception is given by the Kitaev model on the honeycomb lattice~\cite{kitaev2006}, which provides a formidable example for gapless 
and gapped spin liquids. On the experimental side, there has been great development in the search for materials that might be able to support these exotic phases 
of matter. One promising example is given by the so-called Herbertsmithite, which may realize a spin liquid at low temperatures~\cite{norman2016}. Among the variety 
of quantum spin models, the one introduced by Shastry and Sutherland~\cite{shastry1981} deserves particular attention since it gives an example in which the magnetic 
order can be melted by tuning the super-exchange interactions, leading to a particularly simple ground-state wave function, where nearby spins form singlets. Most 
importantly, this Hamiltonian captures the low-temperature properties of SrCu$_2$(BO$_3$)$_2$~\cite{kageyama1999,miyahara1999}. 

\begin{figure}[t!]
\center
\includegraphics[width=0.9\columnwidth]{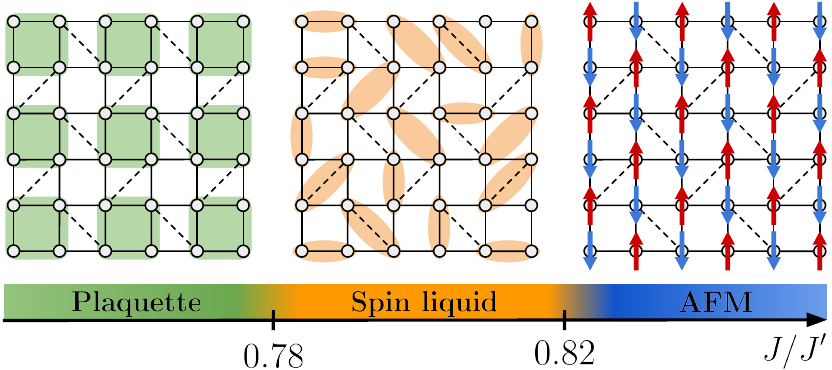}
\caption{\label{fig:phase_diag} 
The ground-state phase diagram of the Shastry-Sutherland model as obtained in this work. The super-exchanges $J$ and $J^\prime$ are denoted by solid and dashed 
lines, respectively.}
\end{figure}

The main interest in this material comes from its properties when external magnetic fields are applied. Indeed, a complicated magnetization curve is observed, with 
various magnetization plateaus (most notably at magnetization $1/8$) that show intriguing properties~\cite{kageyama1999,onizuka2000,kodama2002,corboz2014}. 
The Shastry-Sutherland model is defined by
\begin{equation}\label{eq:SS_hamiltonian}
\hat{{H}} = J \sum_{\langle {\bf r}, {\bf r^\prime} \rangle} \hat{\bf S}_{\bf r} \cdot \hat{\bf S}_{\bf r^\prime}
+ J^\prime \sum_{\langle \langle {\bf r}, {\bf r^\prime} \rangle \rangle} \hat{\bf S}_{\bf r} \cdot \hat{\bf S}_{\bf r^\prime} \, ,
\end{equation}
where $\hat{\bf S}_{\bf r}$ is the $S=1/2$ operator on the site ${\bf r}$ of a $L \times L$ square lattice, with periodic boundary conditions. Here, the first sum 
goes over nearest-neighbor sites on the square lattice, while the second sum is over next-nearest-neighbor sites on orthogonal dimers, according to the bond pattern 
of Fig.~\ref{fig:phase_diag}. For a detailed description of the lattice structure, including its symmetries, see Appendix~\ref{app:symmetries}.

The ground-state properties of the Shastry-Sutherland model are well known in two limiting cases. When $J=0$, the model reduces to a collection of decoupled dimers 
and its ground state is a product of singlets connected by $J^\prime$; this state remains the exact ground state also for finite values of $J/J^\prime$, up to a 
certain value~\cite{shastry1981}. In the opposite limit, when $J^\prime=0$, the Heisenberg model on the square lattice is recovered, whose ground state is the 
N\'eel antiferromagnet; also in this case, the ground state is robust in a finite region when $J^\prime>0$. Despite the substantial effort that has been invested 
in understanding the appearance of magnetization plateaus, the ground-state properties of the Shastry-Sutherland model have been investigated in much less depth. 
One of the first studies based on the mean-field approximation predicted an intermediate helical phase between the dimer and the N\'eel phases~\cite{albrecht1996}, 
while other works suggested a direct transition between these two phases~\cite{miyahara1999,weihong1999}. Later, an intermediate phase with plaquette order has been 
found by series expansion approaches~\cite{koga2000} and confirmed within the generalization to $Sp(2N)$ symmetry and large-$N$ expansion~\cite{chung2001}, by exact 
diagonalizations, and a combination of dimer- and quadrumer-boson methods~\cite{lauchli2002}. Subsequent tensor-network approaches have corroborated the presence of 
the plaquette phase, for $0.675 \lesssim J/J^\prime \lesssim 0.765$~\cite{corboz2013}. This phase breaks the reflection symmetry across the lines containing the
$J^\prime$ bonds (leading to a two-fold degenerate ground state) and is described by resonating singlets on half of the plaquettes with no $J^\prime$ bonds, see 
Fig.~\ref{fig:phase_diag}. The stabilization of plaquette order in SrCu$_2$(BO$_3$)$_2$ has been obtained when hydrostatic pressure is applied, even though there 
is evidence that the broken symmetry is related to the fourfold rotations around the center of plaquettes with no $J^\prime$ bonds~\cite{waki2007,zayed2017}. 
In addition, high-pressure thermodynamics provided evidence of a deconfined quantum critical point between the N\'eel and plaquette phases~\cite{guo2020}. The latter 
aspect has been supported by a numerical analysis, also suggesting the emergence of the $O(4)$ symmetry at the critical point~\cite{lee2019,liu2023}. However, recent 
density-matrix renormalization group (DMRG) and exact diagonalization calculations~\cite{yang2022,wang2022} pushed forward the idea that a spin liquid intrudes 
between the antiferromagnetic and plaquette phases, around $0.79 \lesssim J/J^\prime \lesssim 0.82$. The existence of an intruding spin-liquid phase has been also 
suggested by renormalization group calculations~\cite{kele2022}.

Numerical methods have proven crucial to obtain a description of the physical properties of the Shastry-Sutherland model or, in general, of other complicated 
physical systems. These approaches are mainly based on the variational principle, in which a trial state $\ket{\Psi_{\theta}}$ is introduced, where ${\theta}$ is 
a set of parameters to be optimized in order to minimize the variational energy $\braket{\Psi_{\theta}|\hat{H}|\Psi_{\theta}}/\braket{\Psi_{\theta}|\Psi_{\theta}}$. 
In the variational quantum Monte Carlo scheme~\cite{becca2017}, a quantum system consisting of $N$ spin-1/2 arranged on a lattice is typically studied in the 
computational basis with well-defined spin values along the $z$-axis, i.e., $\{\ket{\sigma} = \ket{\sigma_1^z, \cdots, \sigma_N^z} \}$ with $\sigma_i^z = \pm 1$, 
thus leading to $\ket{\Psi_{\theta}} = \sum_{\{\sigma\}} \Psi_{\theta}(\sigma)\ket{\sigma}$, where $\Psi_{\theta}(\sigma) = \braket{\sigma|\Psi_{\theta}}$ is the 
amplitude of the variational {\it Ansatz}. Different parametrizations of $\Psi_{\theta}(\sigma)$ have been proposed to study frustrated two-dimensional models. 
For example, the description of quantum states able to reproduce the main features of quantum spin liquids is based on the concept of resonating-valence bond 
states~\cite{fazekas1974,anderson1987}, leading to powerful physically inspired wave functions~\cite{ran2007,hu2013,iqbal2016}. Although the construction of this 
kind of wave functions is generalizable to different models, it is not easy to define a systematic way to improve it; as a result, it is not always possible to 
achieve high accuracies for a generic model. On the other hand, DMRG and tensor-network approaches have also proved to be very competitive on two-dimensional 
systems~\cite{yan2011,zhu2015}. Still, despite a great computational effort, two-dimensional systems remain very challenging to deal with. 

In a seminal contribution, Carleo and Troyer~\cite{carleo2017} proposed to parameterize variational states using neural networks, thus defining Neural-Network 
Quantum States (NQS). Further investigations on various many-body systems in one and two spatial dimensions proved that very high accuracies can be obtained with 
this approach~\cite{liang2018,choo2019,sharir2020,hibat2020,viteritti2022,szabo2020,hibat2022,liang2023,chen2023,mezera2023,viteritti2023,rende2023}. Still, in most 
cases their use has been limited to rather simple models, where the exact solutions were already known from other methods (e.g., the unfrustrated Heisenberg model 
on the square lattice or one-dimensional systems)~\cite{carleo2017,sharir2020,hibat2020,viteritti2022}. Attempts to address challenging cases have been pursued, but
without addressing important open questions on the ground-state properties~\cite{liang2018,choo2019,szabo2020,hibat2022,liang2023,chen2023,mezera2023}. In addition, 
neural-network architectures have also been employed to enhance conventional variational states, which were widely utilized in previous studies on frustrated spin 
models (e.g., Gutzwiller-projected fermionic states)~\cite{nomura2017,ferrari2019,nomuraimada2021}. Moreover, NQS are particularly promising to resolve challenging 
problems in strongly-correlated systems, since they can efficiently represent highly-entangled quantum states~\cite{levine2019,sharir2022}. On the contrary, DMRG 
and related Tensor Network approaches can accurately describe states with high entanglement only in one-dimensional systems, where a large bond dimension can be 
easily used. Instead, in two dimensions, serious limitations appear, either imposing to work with a high-rank tensor structure or a quasi-one-dimensional cluster 
(with low-rank tensors arranged in a snaked path~\cite{stoudenmire2012}).

\begin{figure*}[t]
\includegraphics[width=2\columnwidth]{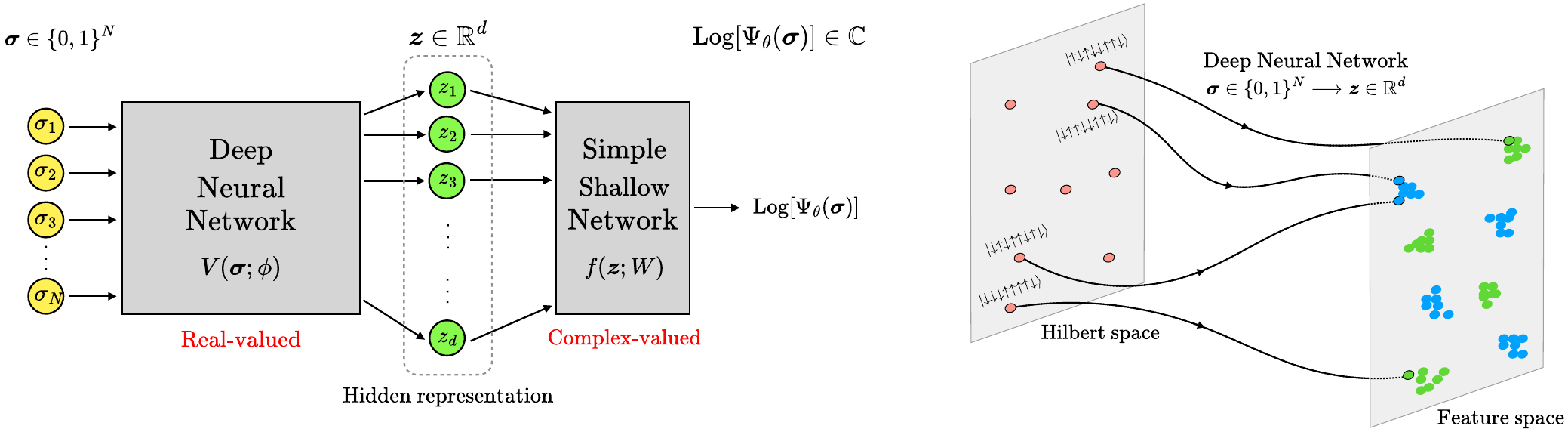}
\caption{\label{fig:realcomplex} 
\textbf{Left panel:} The NQS is defined as the composition of two functions: first, a deep neural network $V(\boldsymbol{\sigma}; \phi)$ (with real-valued 
parameters) maps the input configurations $\boldsymbol{\sigma}$ into hidden representations $\boldsymbol{z}$; then, a simple shallow network $f(\boldsymbol{z}; W)$ 
(with complex-valued parameters) generates the logarithm of the amplitudes $\text{Log}[\Psi_{\theta}(\boldsymbol{\sigma})]$ starting from hidden representations.  
\textbf{Right panel:} Pictorial illustration of the mapping process carried out by the deep neural network. The network maps spin configurations from the Hilbert 
space $\boldsymbol{\sigma}$ into configurations in a feature space $\boldsymbol{z} \in \mathbb{R}^d$ with a non-trivial structure. In subsection~\ref{subsec:hidden_rep},
we show for example that physical configurations $\boldsymbol{\sigma}$ cluster in feature space according to the sign of the amplitudes 
$\text{Log}[\Psi_{\theta}(\boldsymbol{\sigma})]$.}
\end{figure*}

Here, we aim to push the boundaries of this approach by demonstrating that an {\it Ansatz} exclusively reliant on neural networks enables us to achieve unprecedented 
accuracy in solving the challenging Shastry-Sutherland model. This model poses a particularly demanding problem in the realm of highly-frustrated magnetism, and our 
approach facilitates the extraction of its intricate physical properties. Specifically, we use an architecture based on Transformer~\cite{vaswani2017,dosovitskiy2021} 
which has already proven to be extremely accurate for models in one and two dimensions~\cite{viteritti2023,rende2023,diluo2023,czischek2023}. However, in this work, 
we incorporate the Transformer architecture in an innovative framework where the deep neural network is employed as a map from the space of the physical spin 
configurations to an abstract space, where the determination of the low-energy properties of the systems is simplified. This approach mirrors the {\it representation 
learning} that is central to the success of modern deep learning~\cite{bengio2014}. Carrying out simulations on clusters with periodic-boundary conditions, we show 
that there exists a small, but finite, region in the phase diagram in which both the antiferromagnetic and plaquette order parameters vanish in the thermodynamic 
limit (see Fig.~\ref{fig:phase_diag}). As a result, this region is consistent with the existence of a spin-liquid state. Another original contribution of this work 
is to define a suitable modification of the ViT architecture to treat excited states at finite momenta. This approach lends support to the existence of a {\it gapless} 
spin liquid.

\section{The variational wave function}\label{sec:wavefunction}

In this study, we take a new perspective on NQS by leveraging the principle of {\it representation learning}~\cite{bengio2014} that is key to the success of deep 
neural networks in practice. The idea is that the mathematical structure of deep networks, a composition of simple functions with parameters that can be tuned to 
data, allows neural networks to automatically extract the pertinent features of a data set for a given task. These features or {\it representations} of the inputs 
are then used for downstream tasks, like, in our case, predicting the amplitude of the wave function for a given spin configuration. The idea of {\it learning} 
these representations directly from data is contrary to the approach of classical machine learning, which required careful engineering and considerable domain 
expertise to distil raw data (such as the spin configurations) into a representation or feature vector that could be used for a downstream task~\citep{lecun2015deep}.

Here, we follow this approach by building a variational {\it Ansatz} where we use a deep neural network to map physical spin configurations into a feature space. 
This transformation enables an accurate prediction of the amplitude associated with each configuration with even a simple, shallow fully-connected 
layer~\cite{carleo2017}. By reframing the NQS as feature extractors rather than just universal approximators of complicated functions, the variational state is 
naturally perceived as the composition of two distinct functions, each with a specific role:
\begin{equation}\label{eq:composition}
\begin{aligned}
    \boldsymbol{z} &= V(\boldsymbol{\sigma}; \phi) \ , \\
    \text{Log}[\Psi_{\theta}(\boldsymbol{\sigma})] &= f \left( \boldsymbol{z}; W \right)\ ,
\end{aligned}
\end{equation}
where the variational parameters are partitioned into two blocks $\theta = \{\phi, W\}$. The function $V(\boldsymbol{\cdot}; \phi)$ is parameterized as a {\it deep} 
neural network, mapping physical configurations $\boldsymbol{\sigma}$ to vectors $\boldsymbol{z}$, called {\it hidden representations}, which belong to a 
$d$-dimensional {\it feature space}. Conversely, $f(\boldsymbol{\cdot}; W)$ is a {\it shallow} neural network used to generate a single scalar value from the 
hidden representations $\boldsymbol{z}$. This final value is used to predict the amplitude corresponding to the input configuration. In order to predict both 
modulus and phase of the variational state (which is fundamental in cases where the exact sign is not known {\it a priori}), it is convenient to employ a 
complex-valued variational state. The structure of the {\it Ansatz} in Eq.~\eqref{eq:composition} suggests the possibility of taking $\phi$ as real-valued 
parameters in the deep neural network $V(\boldsymbol{\cdot}; \phi)$. Subsequently, only the parameters $W$ of the shallow function $f(\boldsymbol{\cdot}; W)$ 
can be taken complex-valued. We schematically represent these two steps in the left panel of Fig.~\ref{fig:realcomplex}; instead, a pictorial scheme of the mapping 
process from the physical space of the spin configurations to the feature space is depicted in the right panel of Fig.~\ref{fig:realcomplex}.

Far from being only a change of viewpoint, the possibility of having a real-valued feature extractor is crucial in practice. Several works showed recently that 
depth is crucial to achieve high accuracies on two-dimensional quantum systems~\cite{chen2023,li2022,liang2022,roth2023}. However, training deep networks is a 
complicated task that is only possible by leveraging techniques such as Layer Normalization~\cite{xiong2020}, skip connections~\cite{he2015deep}, and appropriate 
activation functions~\cite{agarap2019deep}. However, all of these techniques have been developed for real-valued architectures, and cannot be straightforwardly
generalized to complex-valued neural networks. For these reasons, in Ref.~\cite{viteritti2023}, the optimization of a deep Transformer architecture having 
complex-valued parameters necessitated the development of a heuristic procedure involving the introduction of a cut in the attention weights. A big advantage of 
the newly proposed {\it Ansatz} with a real-valued feature extractor is then that it can be trained from scratch without additional restrictions and with minimal 
regularization in the optimization protocol (see Appendix~\ref{app:optimization} for details). This modified architecture has recently yielded state-of-the-art
results on one of the most popular benchmark in frustrated magnetism~\cite{rende2023}. The following two subsections give a detailed description of the architecture 
of the neural network that we use to study the Shastry-Sutherland model; we present our results in section~\ref{sec:results}.

\begin{figure}[t]
\includegraphics[width=0.9\linewidth]{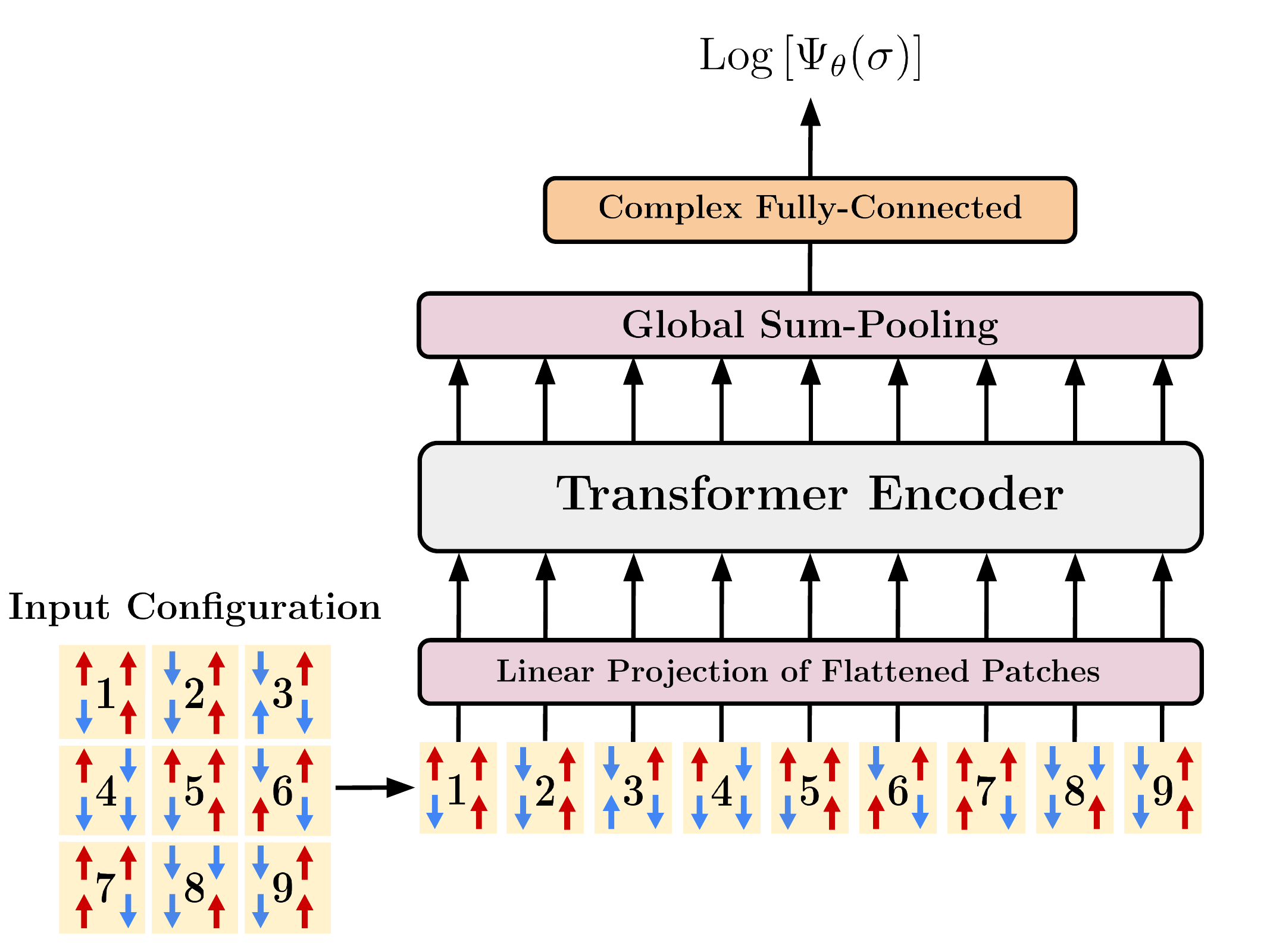}
\caption{\label{fig:architecture}
The input spin configuration $\boldsymbol{\sigma}$ is partitioned into patches, which are linearly projected in a $d$-dimensional embedding space and then processed 
by a Vision Transformer. The latter one builds new representations of the patches, which are then combined through summation and fed into a final single complex-valued
fully-connected layer in order to obtain the logarithm of the (complex) wave function. Notice that this is a particular instantiation of the more general scheme 
proposed in the left panel of Fig.~\ref{fig:realcomplex}.}
\end{figure}

\begin{figure}[t]
\includegraphics[width=0.4\columnwidth]{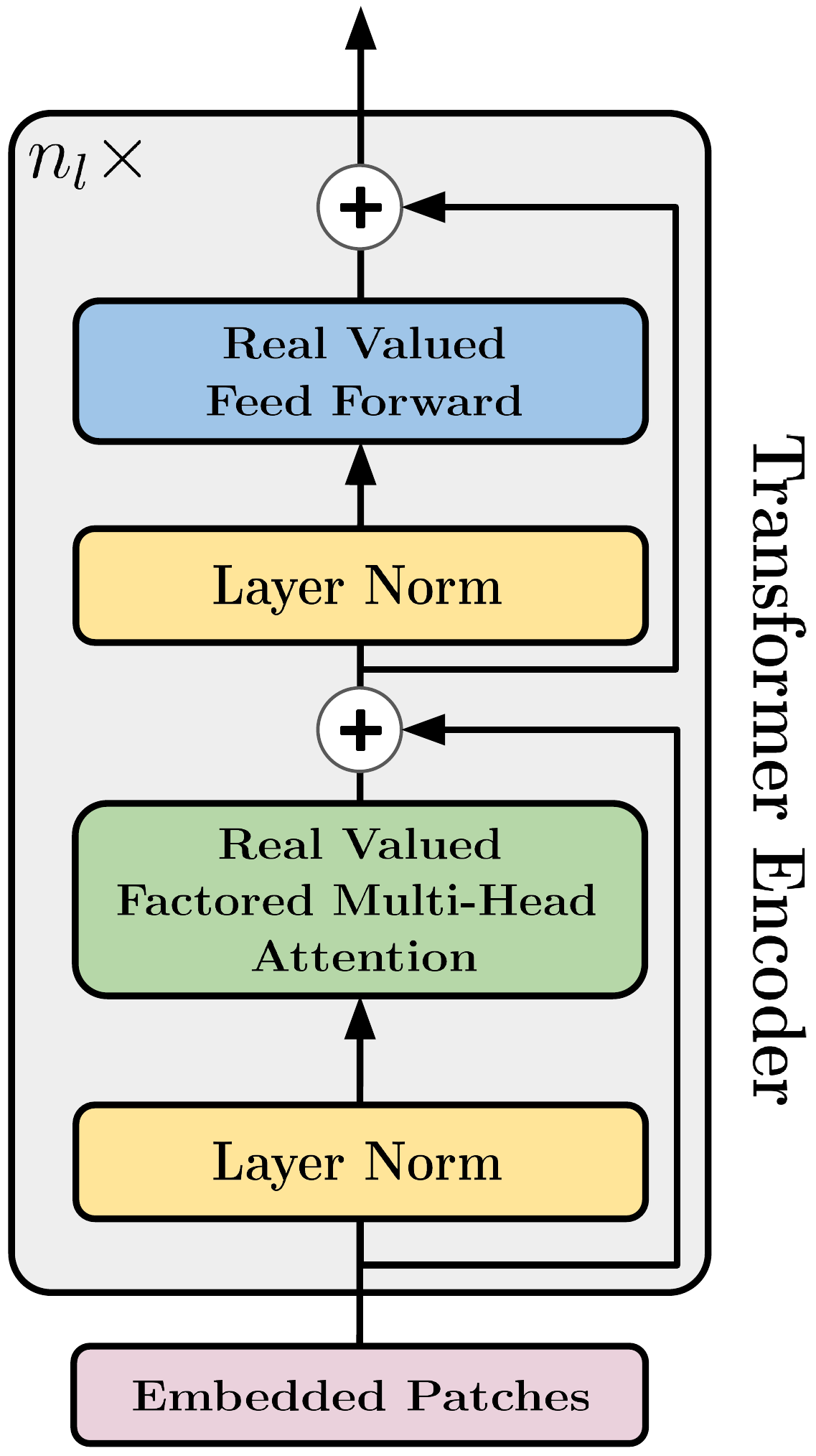}
\caption{\label{fig:encoder} 
To process the embedded patches, each Transformer Encoder block employs a real-valued factored multi-head attention mechanism, which mixes the patches, and a 
real-valued two-layers Feed-Forward neural network, which is used to introduce a non-linearity. Skip connections and Layer Normalization are also employed.}
\end{figure}

\begin{figure*}[ht!]
\includegraphics[width=2.05\columnwidth]{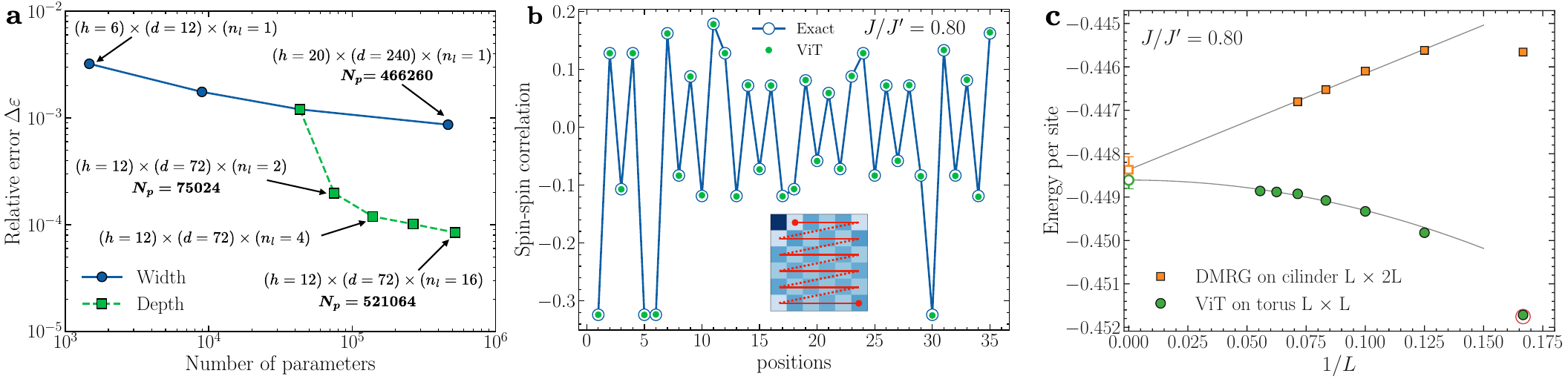}
\caption{\label{fig:benchmarks} 
\textbf{Panel a:} Relative error $\Delta\varepsilon = |(E_{\text{exact}} - E_{\text{ViT}}) / E_{\text{exact}}|$ of the ViT wave function on a $6\times 6$ lattice at 
$J/J^\prime=0.8$. First, fixing only one layer and measuring the accuracy by increasing the width (blue dots). Then, for a fixed width, by increasing the number of 
layers (green squares). \textbf{Panel b:} The isotropic spin-spin correlations in real space as computed by the ViT wave function (full dots) on a $6\times 6$ lattice 
at $J/J^\prime=0.8$. Values from exact diagonalization (empty dots) are also shown for comparison. Inset: The red line shows how the spin-spin correlations are ordered 
in the panel (b). \textbf{Panel c:} The comparison between the energies per site obtained by the ViT wave function (green circles) on $L\times L$ lattices with 
periodic-boundary conditions from $L=6$ to $L=18$ and the ones obtained by DMRG (orange squares) on $2L\times L$ cylinders with open-boundary conditions along the 
$x$ direction from $L=6$ to $L=14$~\cite{yang2022}. The exact result on the $6\times 6$ lattice is denoted with an empty red circle. The reported energy values were 
obtained by optimizing a ViT model with hyperparameters set to $h=12$, $d=72$, and $n_l=8$, utilizing a sample size of $M=6 \times 10^3$ during the optimization 
process.}
\end{figure*}

\subsection{Vision Transformer}\label{sec:vit}

One of the most promising architectures in machine-learning applications is the Transformer~\cite{vaswani2017}, which, originally designed for natural language 
processing tasks, rapidly reached competitive results also in different fields, for example the Vision Transformer (ViT) for image classification 
tasks~\cite{dosovitskiy2021}. Some of us adapted the ViT architecture to study one-dimensional systems~\cite{viteritti2023}, achieving results that are comparable 
with DMRG on large clusters. In this work, we propose its use to parametrize $V(\boldsymbol{\cdot}; \phi)$ in Eq.~\eqref{eq:composition}, instead the function $f$ 
is chosen to be:
\begin{equation}\label{eq:rbm}
    f(\boldsymbol{z}; W) = \sum_{\alpha=1}^K \log\cosh \left( b_{\alpha} + \boldsymbol{w}_{\alpha} \cdot \boldsymbol{z} \right) \ ,
\end{equation}
where the variational parameters $W$ are the biases and the weights of the linear transformation. The number of hidden neurons $K$ is a hyperparameter of the 
network. Notice that Eq.~\eqref{eq:rbm} has the same functional form as the well-known Restricted-Boltzmann Machine (RBM) introduced by~\citet{carleo2017}. 
Crucially, in this case it is not applied to the physical configuration $\boldsymbol{\sigma}$ but instead to the hidden representation $\boldsymbol{z}$. This 
is the change of paradigm that we want to emphasize. With these choices, the process of constructing the amplitude corresponding to a physical spin configuration 
$\boldsymbol{\sigma}$ involves the following steps (see Fig.~\ref{fig:architecture}):
\begin{enumerate}
\item The input spin configuration $\boldsymbol{\sigma}$ is initially divided into $n$ patches (see Appendix~\ref{app:symmetries}).
\item The patches are linearly projected into a $d$-dimensional embedding space, resulting in a sequence of vectors $(\mathbf{x}_1, \cdots, \mathbf{x}_n)$, where 
$\mathbf{x}_i \in \mathbb{R}^d$.
\item A ViT processes these embedded patches, producing another sequence of vectors $(\mathbf{y}_1, \cdots, \mathbf{y}_n)$, where $\mathbf{y}_i \in \mathbb{R}^d$.
\item The hidden representation $\boldsymbol{z}$ of the configuration $\boldsymbol{\sigma}$ is defined by summing all these output vectors: 
$\boldsymbol{z}=\sum_{i=1}^n \mathbf{y}_i$.
\item A fully-connected layer with complex-valued parameters, defined in Eq.~\eqref{eq:rbm}, produces the amplitude $\text{Log}[\Psi_{\theta}(\boldsymbol{\sigma})]$ 
corresponding to the input configuration $\boldsymbol{\sigma}$. Specifically, we set $K=d$.
\end{enumerate}

Notably, while the vector $\mathbf{x}_i$ depends solely on the spins contained in the $i$-th patch, the resulting vector $\mathbf{y}_i$ is a function of all the 
spins in the configuration. The ViT architecture is constructed as a sequence of $n_l$ encoder blocks. In each of them, a multi-head self-attention layer (with 
$h$ heads) is followed by a two-layer fully connected network. For a detailed description of the Encoder Block see subsection~\ref{subsec:encoder_block}.

Notice that the structure of this variational {\it Ansatz} requires a large number of parameters. In order to optimize them, modern formulations of the Stochastic 
Reconfiguration technique~\cite{sorella2005}, able to deal with a large number of variational parameters~\cite{chen2023,rende2023}, are used (see 
Appendix~\ref{app:optimization}).

\subsection{Encoder Block}\label{subsec:encoder_block}

The Encoder Block is the core of the Transformer architecture (see Fig.~\ref{fig:encoder}). The input sequence of the $l$-th Encoder Block (where $l$ runs from 1 to
$n_l$) is the set of $n$ vectors $(\mathbf{x}_1, \cdots, \mathbf{x}_n)$, where, for the sake of simplicity, the index $l$ is not made explicit. This sequence of 
vectors is processed by a real-valued factored multi-head attention mechanism~\cite{rende2023b,queries_keys}. The $\mu$-th attention vector $\mathbf{A}_i^\mu$ is 
defined by first applying a {\it local} linear transformation $\boldsymbol{V}^\mu$ to each input vector $\boldsymbol{x}_j$. 

\begin{figure*}[ht!]
\includegraphics[width=1.7\columnwidth]{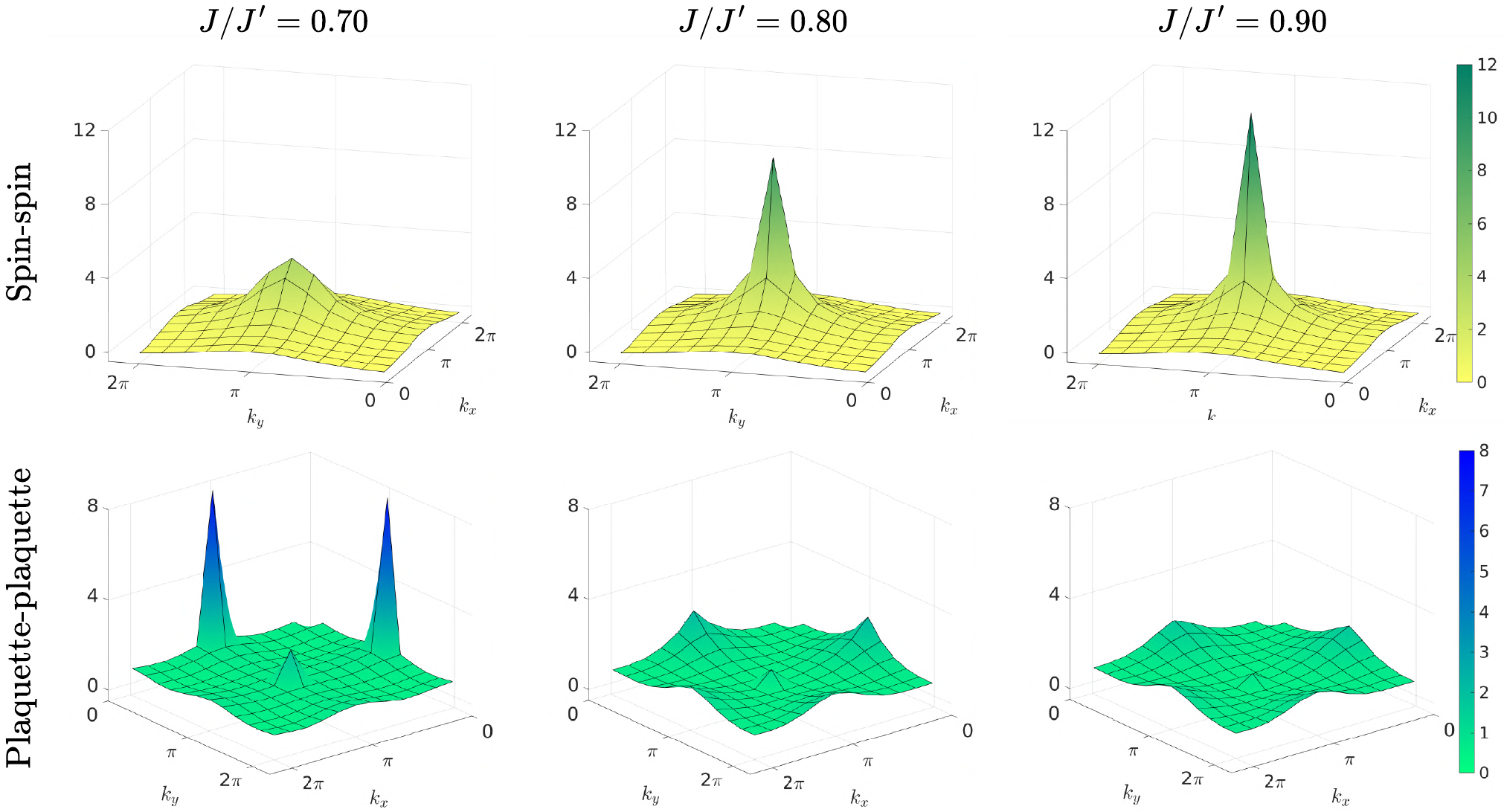}
\caption{\label{fig:structure_factor} 
Fourier transform of the spin-spin (upper panels) and plaquette-plaquette (lower panels) correlations for $L=12$ for different values of the frustration ratio 
$J/J^\prime$. The calculations are performed with a Vision Transformer characterized by a number of heads equal to $h=12$, an embedding dimension $d=72$, and number 
of layers $n_l=8$.}
\end{figure*}

The resulting vectors $V^\mu \mathbf{x}_j$ are then {\it globally} mixed according to the attention mechanism~\cite{vaswani2017}
\begin{equation}\label{eq:fact_attention}
    \mathbf{A}_i^\mu = \sum_j \alpha_{i-j}^\mu V^\mu \mathbf{x}_j,
\end{equation}
where $\mu=1, \dots, h$, with $h$ the numbers of heads in the multi-head attention mechanism. The parameters $\alpha_{i-j}^\mu \in \mathbb{R}$ are the attention 
weights, which define the so-called {\it attention maps} (see subsection~\ref{subsec:att_maps}). The $h$ different attention representations computed in each head 
$\mathbf{A}_i^\mu \in \mathbb{R}^{d/h}$ are concatenated together to give an output sequence of $n$ attention vectors $(\mathbf{A}_1, \cdots, \mathbf{A}_n)$, 
with $\mathbf{A}_i \in \mathbb{R}^d$. Then, after another linear projection which mixes the representations of the different heads, each attention vector is finally 
passed identically and independently through a non-linearity, which is taken to be a (real-valued) two-layers fully-connected neural network, with hidden dimension 
$2d$ and the standard rectified linear unit (ReLU) activation function. The output of the $l$-th encoder block is a sequence $(\mathbf{y}_1, \cdots, \mathbf{y}_n)$, 
with $\mathbf{y}_i \in \mathbb{R}^d$ being a new representation of the $i$-th input. Pre-Layer Normalization~\cite{xiong2020} and skip connections are used, these 
being the key elements that permit the optimization of deep networks. The use of factored attention in Eq.~\eqref{eq:fact_attention} is justified by the physical 
interpretation we give to the attention weights; indeed, we expect that they should mainly depend on the relative positions among groups of spins and not on the 
actual values of the spins in the patches~\cite{viteritti2023,queries_keys}. Moreover, attention weights are taken translationally invariant, in order to encode 
the translational symmetry between patches.

\section{Results}\label{sec:results}

We consider $L \times L$ clusters in which sites ${\bf r}=a/2(x,y)$ are labelled by $x$ and $y$ that take values from $0$ to ${L-1}$, and periodic-boundary 
conditions along the primitive vectors ${\bf a}_1=(a,0)$ and ${\bf a}_2=(0,a)$, with $a=2$. As a result, the total number of sites is $L^2$, while the number of 
unit cells is $L^2/4$. Most of the calculations are performed on clusters with $L$ ranging from $6$ to $14$; in addition, calculations with $L=16$ and $18$ have 
been considered for $J/J^\prime=0.8$, located in the middle of the putative spin-liquid region. We mention that the ViT wave function breaks the spin $SU(2)$ 
symmetry and, therefore, it is not an eigenstate of the total spin $S^2$; still, it is possible to fix the $z$-component of the total spin, by performing the 
Monte Carlo sampling within a sector with a fixed value of $S^z=1/2 \sum_i \sigma_i^z$. All calculations for assessing ground-state properties are performed 
taking $S^z=0$ (see subsections~\ref{subsec:benchmarks} and~\ref{subsec:phase_diag}). By contrast, triplet excitations are obtained by choosing $S^z=1$ (see 
subsection~\ref{subsec:spinliquid}). We emphasize that, the optimized states have extremely small deviations from the expected value of the total spin, e.g., 
$S(S+1) \approx 0.002$ ($0.1$) and $2.002$ ($2.07$) for the singlet and triplet states for $L=6$ ($L=10$) cluster, respectively.

\subsection{Benchmarks}\label{subsec:benchmarks}

In order to validate our approach, we compare the results obtained by the ViT wave function with those obtained by exact diagonalizations on a small $6 \times 6$ 
cluster. Specifically, we focus on the challenging point $J/J^\prime=0.8$. We first examine the accuracy of the variational energies while varying the hyperparameters 
of the neural network. In Fig.~\ref{fig:benchmarks}a, we present the relative energy error as a function of the number of parameters, distributed in two different 
ways within the architecture. Initially, we maintain a single layer ($n_l=1$) and increase the number of heads $h$ and embedding dimension $d$. Subsequently, we 
fix a specific width ($h=12$ and $d=72$) and increment the number of layers from $n_l=2$ to $n_l=16$ (the energies are reported in Appendix~\ref{app:extrapol}). 

\begin{figure*}[ht!]
\includegraphics[width=2\columnwidth]{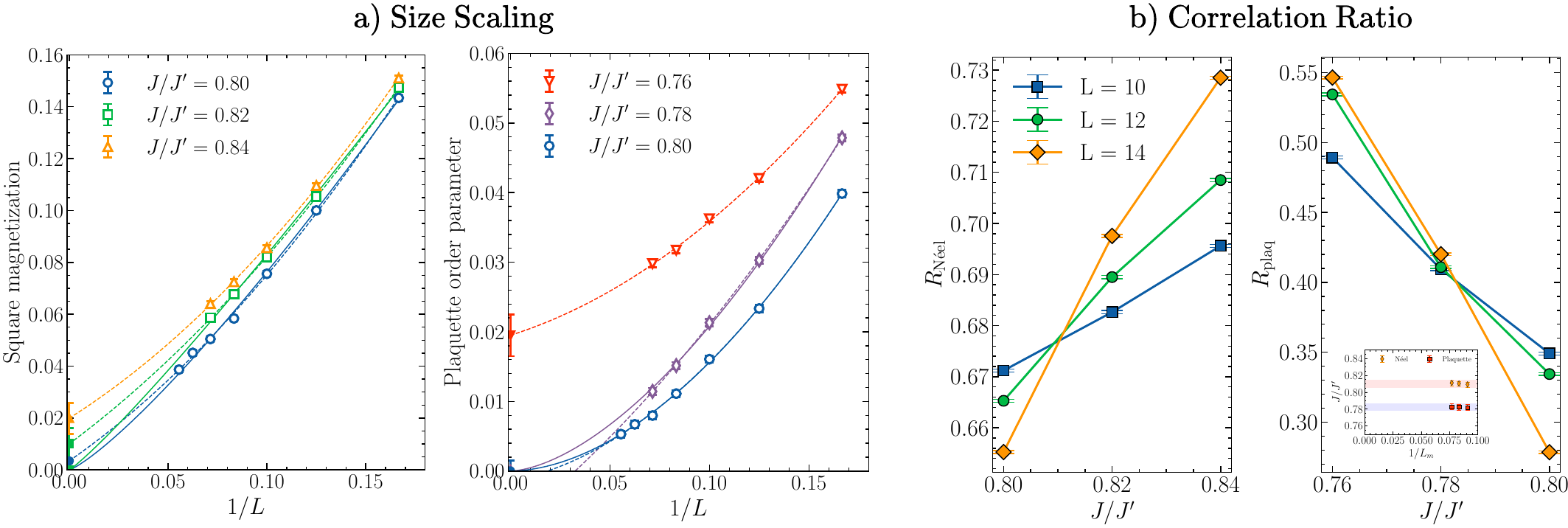}
\caption{\label{fig:phase_transition}
\textbf{Panel a:} Size scaling of the square magnetization $m^2(L)$ (left panel), and the plaquette order parameter $m_p(L)$ (right panel) as a function of $1/L$ 
from $L=6$ up to $L=18$. The values reported for each size $L$ are obtained by extrapolating to an infinite number of layers (see Appendix~\ref{app:extrapol} for the
details), except for $L=16$ and $L=18$, where only simulations with $n_l=8$ layers have been performed. The error bars of the extrapolated values in the thermodynamic 
limit are estimated via a resampling technique with gaussian noise. The fits associated to dashed curves are obtained using second-order polynomials in $1/L$ (see 
Eq.~\eqref{eq:quadratic_fitting} of the main text), while solid curves are obtained using the critical form in Eq.~\eqref{eq:critical_fitting} of the main text. 
\textbf{Panel b:} In the left (right) panel we show the correlation ratio $R_{\text{N\'eel}}$ ($R_{\text{plaq}}$) for the antiferromagnetic (plaquette) order in 
the interval $J/J^\prime \in [0.80, 0.84]$ ($J/J^\prime \in [0.76, 0.80]$). System sizes from $L=10$ to $L=14$ are used, and correlation ratio values are computed 
exclusively with architectures having $n_l=8$ layers. Inset: Crossing points of the correlation ratio for N\'eel (orange diamond) and plaquette (red squares) order 
parameter as a function of the system size.  The crossing points are obtained using $L_1 \times L_1$ and $L_2 \times L_2$ clusters with 
$(L_1, L_2) = (10,12), (10, 14), (12,14)$, with $L_m=(L_1+L_2)/2$. Error bars on the correlation ratio are determined using resampling techniques under the assumption 
of Gaussian noise.}
\end{figure*}

This analysis highlights the importance of the model depth: for a fixed number of parameters, architectures that allocate parameters across multiple layers exhibit 
superior accuracy. These results align with previous works~\cite{chen2023,rende2023,li2022,liang2022,roth2023}, which underscore the necessity of deep neural 
networks for achieving high-precision results in two-dimensional frustrated systems. In addition, in Fig.~\ref{fig:benchmarks}b we show the isotropic spin-spin 
correlation functions $\braket{\hat{\bf S}_{\bf 0}\cdot \hat{\bf S}_{\bf r}}$, illustrating that our variational wave function not only yields accurate energies, 
but also faithfully captures correlation functions at all distances. For cluster sizes exceeding $L=6$, exact results become unattainable. Consequently, in 
Fig.~\ref{fig:benchmarks}c, we compare the variational energies of the ViT {\it Ansatz} on $L\times L$ clusters (with periodic-boundary conditions) to the ones 
obtained using the DMRG method on $L_x \times L_y$ cylinders with open and periodic boundaries in the $x$ and $y$ direction, respectively ($L_x=2L_y$ and $L_y=L$ 
are considered)~\cite{yang2022}. The energy per site is extrapolated in the thermodynamic limit, using system sizes ranging from $L=8$ to $L=18$ for the ViT wave 
function, and from $L=8 $ to $ L=14$ for the DMRG. To enhance the efficiency of the ViT for larger systems, specifically at $L=16$ and $L=18$, we employ a {\it local} 
attention mechanism (see Appendix~\ref{app:large_lattices} for further details). We mention that the energies obtained by the ViT wave function reveal a $1/L^2$ 
term as the leading correction, whereas the DMRG results exhibit an additional $1/L$ term. Most importantly, the energy extrapolated in the thermodynamic limit 
is compatible within the two approaches.

\begin{figure*}[t]
\includegraphics[width=2\columnwidth]{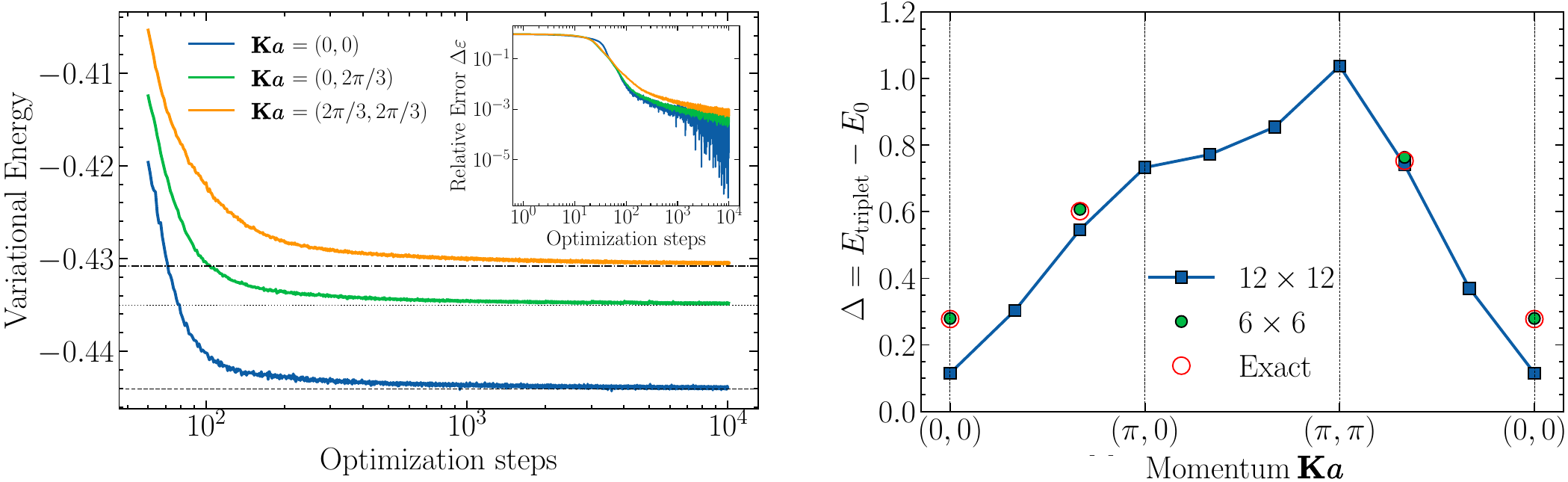}
\caption{\label{fig:excited} 
\textbf{Left panel:} Variational energy as a function of the optimization steps for the triplet momentum-resolved excitation spectra on a $6 \times 6$ lattice, 
with ${\bf K}a = (0,0)$, $(0, 2\pi/3)$, and $(2\pi/3, 2\pi/3)$; depicted in blue, green, and orange, respectively. The exact energies per site for the corresponding 
states are denoted by dashed ($E=-0.444040$), dotted ($E=-0.435033$), and dash-dotted ($E=-0.430831$) lines. The inset displays the relative error of the variational 
energy compared to exact diagonalization results. \textbf{Right panel:} The energy gap $\Delta$ between the ground state and the triplet excitations for the 
independent momenta of $6 \times 6$ lattice (green points), with the corresponding values obtained from exact diagonalization (red empty points). The energy gaps 
for a $12 \times 12$ lattice on a closed path in the Brillouin zone are denoted by blue squares. All the calculations are performed at a frustration ratio of 
$J/J^\prime=0.8$. The hyperparameters for these calculations are $h=12$, $d=72$, and $n_l=8$. For the optimization protocol, we used a sample size of $M=2^{14}$, 
a learning rate $\tau=0.02$, and a diagonal shift regularization term of $\lambda = 10^{-3}$ (see Appendix~\ref{app:optimization}).}
\end{figure*}

\subsection{Phase diagram}\label{subsec:phase_diag}

Having proved the high accuracy of our {\it Ansatz}, we now focus on the region $0.7 \le J/J^\prime \le 0.9$, which is expected to include both antiferromagnetic 
and plaquette phases, as well as the putative spin-liquid one. The presence of antiferromagnetic order is extracted from the thermodynamic limit of the staggered 
magnetization $m^2(L) = S(\pi,\pi)/L^2$~\cite{yang2022}, where 
\begin{equation} 
S({\bf k})= \sum_{\bf r} e^{i{\bf k}\cdot {\bf r}} \braket{\hat{\bf S}_{\bf 0} \cdot \hat{\bf S}_{\bf r}} 
\end{equation}
is the spin structure factor. Since the antiferromagnetic order pertains to the square lattice denoted by the sites ${\bf r}$, i.e., {\it without} considering the 
basis of the Shastry-Sutherland lattice, it is useful to define the momenta within this convention, i.e., ${\bf k}=2\pi/L(n,m)$ with $n$ and $m$ taking values
from $0$ to $L-1$. The existence of N\'eel order is signalled by a diverging peak at ${\bf k_m}=(\pi,\pi)$. In addition, the insurgence of the plaquette order is 
detected by a suitably defined order parameter 
\begin{equation}\label{eq:plaq_order}
m_p(L)=|C(L/2,L/2)-C(L/2-1,L/2-1)| \ , 
\end{equation}
where the function $C({\bf r})$ is defined as follows: starting from the operator $\hat{P}_{\bf r}$, which performs a cyclic permutation of the four spins of a 
plaquette with the top-right site at ${\bf r}$~\cite{yang2022}, the following correlation functions are evaluated:
\begin{equation}\label{eq:corr_plaq}
C({\bf r}) = \frac{1}{4} \braket{[\hat{P}_{\bf r} + \hat{P}^{-1}_{\bf r}][\hat{P}_{\bf 0} + \hat{P}^{-1}_{\bf 0}]} \ .
\end{equation}
Therefore, the plaquette order parameter $m_p(L)$ of Eq.~\eqref{eq:plaq_order} measures the difference, along the diagonal, of the plaquette correlation at the 
maximum distance and the second maximum distance; whenever the plaquette order is present, the correlation along the diagonal does not decay to zero, implying a 
non-vanishing value of $m_p(L)$ for large $L$. Similarly, the Fourier transform of the correlation functions in Eq.~\eqref{eq:corr_plaq} (with the same conventions 
as for spins) denoted by $C({\bf k})$ can be analyzed. The presence of the plaquette order can be identified by a diverging peak at ${\bf k_p}=(0,\pi)$ or $(\pi,0)$. 
The results for $L=12$ are shown in Fig.~\ref{fig:structure_factor}, for three values of the frustration ratio: for $J/J^\prime=0.7$ the ground state has strong 
peaks in $C({\bf k})$ and a rather smooth spin structure factor $S({\bf k})$, which is typical of a state with plaquette order; by contrast, for $J/J^\prime=0.9$ 
there are strong spin-spin correlations and weak plaquette-plaquette ones, which is characteristic of antiferromagnetic states.

In between, for $J/J^\prime=0.8$, the spin-spin correlations still have a peak, with moderate plaquette correlations. In order to get information on the thermodynamic 
limit, a size scaling is necessary. In general, if magnetic order is stabilized, the square magnetization scales asymptotically as~\cite{sandvik1997,calandra1998}:
\begin{equation}\label{eq:quadratic_fitting}
    m^2(L) \approx m_0^2 + \frac{A_1}{L} + \frac{A_2}{L^2} \ ,
\end{equation}
where $m_0$ is the magnetization in the thermodynamic limit. In a disordered phase, the magnetization vanishes in the thermodynamic limit. The size corrections can 
be either exponential (for a gapped state) or power law (for a gapless one). In the vicinity of the N\'eel transition, the gap is relatively small and we use the 
``critical'' form~\cite{nomuraimada2021}:
\begin{equation}\label{eq:critical_fitting}
    m^2(L) \approx L^{-(1+\eta)} \ .
\end{equation}
Similar scaling behaviors are considered for $m_p(L)$ (within the plaquette phase, exponential corrections should be present, but no appreciable differences in the
fits are observed with respect to the choice of a polynomial fit). In Fig.~\ref{fig:phase_transition}a, we perform a size-scaling extrapolation of both order parameters.
For $J^\prime/J = 0.84$ ($J^\prime/J = 0.76$), the numerical values of the square magnetization (plaquette order parameter) fit well with a second-order polynomial
in $1/L$ and suggest the existence of long-range order in the thermodynamic limit. By contrast, for $J^\prime/J=0.78, 0.8$ ($J^\prime/J=0.82, 0.84$), a more 
appropriate description of the scaling behavior of $m^2$ ($m_p$) is obtained by the critical relation of Eq.~\eqref{eq:critical_fitting} (more details about the
extrapolations are reported in the Appendix~\ref{app:extrapol}). This fact is compatible with the existence of a {\it gapless} spin liquid, which is also corroborated
by the direct computation of the spin gap (see subsection~\ref{subsec:spinliquid}). Interestingly, fitting the data of the square magnetization at $J^\prime/J=0.8$ 
with $m^2 \approx L^{-(1+\eta)}$, we get $\eta \approx 0.3$, in agreement with the DMRG calculations of Ref.~\cite{yang2022}. We emphasize that, for the most 
challenging point $J/J^\prime=0.8$, lattices with $L=16$ and $18$ have been also considered, giving further support in favor of an intermediate spin liquid phase.

In summary, we find that the magnetization (plaquette order) vanishes for $J/J^\prime \approx 0.82$ ($J/J^\prime \approx 0.77$). These results suggest that a 
spin liquid exists between ${(J/J^\prime)_{\text{plaq}} \approx 0.77}$ and ${(J/J^\prime)_{\text{N\'eel}}\approx 0.82}$. To further support the present outcome, 
we measure the correlation ratio for the plaquette order as ${R_{\text{plaq}}=1-C(\bf{k_p}+\delta \bf{k})/C(\bf{k_p})}$, and for the magnetic order as 
${R_{\text{N\'eel}}=1-S(\bf{k_m}+\delta \bf{k})/S(\bf{k_m})}$, where ${||\delta {\bf k}||=2\pi/L}$. When plaquette (magnetic) order is not present, 
$C(\bf{k})$ ($S(\bf{k})$) is a smooth function of $\bf{k}$, which implies that $R_{\text{plaq}} \to 0$ ($R_{\text{N\'eel}} \to 0$) in the thermodynamic limit; 
instead, when plaquette (magnetic) order settles down, $C(\bf{k})$ ($S(\bf{k})$) is finite for all the momenta except for $\bf{k_p}$ ($\bf{k_m}$), leading to 
$R_{\text{plaq}} \to 1$ $(R_{\text{N\'eel}} \to 1)$. Then, the transition point may be accurately determined by locating the crossing point of the correlation 
ratio curves for different system sizes. The results for the plaquette (magnetic) order are shown in Fig.~\ref{fig:phase_transition}b, in the relevant interval 
$J/J^\prime\in [0.76,0.80]$ ($J/J^\prime\in [0.80,0.84]$), increasing the system size, i.e., for $L=10$, $12$, and $14$. The various curves cross at 
$(J/J^\prime)_{\text{plaq}}\approx 0.78$ ($(J/J^\prime)_{\text{Néel}}\approx 0.81$), validating the phase boundary derived from the extrapolations of the order 
parameters.

\subsection{Nature of the Spin Liquid}\label{subsec:spinliquid}

A crucial step toward understanding the nature of the spin liquid-phase is the characterization of its low-energy excitations. The first important question is to
determine whether the energy spectrum is gapped or gapless. Here, we focus on the frustration ratio $J/J^\prime=0.8$, a representative point within the spin-liquid 
phase, and assess triplet excitations for different momenta. Then, we focus on the lowest-energy state, which lies at the $\Gamma$ point of the Brillouin zone, and
perform the extrapolation to the thermodynamic limit. Our results provide strong evidence for a {\it gapless} spin liquid.

The ViT architecture outlined in Section~\ref{sec:vit} employs translationally invariant attention weights and the input patches are constructed from spins within 
the unit cell (see Appendix~\ref{app:symmetries} for additional details). Consequently, the resulting wave function preserves translational symmetry with zero 
momentum. Nevertheless, by exploiting the translational equivariance property of the mapping between the input vectors $(\boldsymbol{x}_1, \dots, \boldsymbol{x}_n)$ 
and output vectors $(\boldsymbol{y}_1, \dots, \boldsymbol{y}_n)$, the architecture can be easily adapted to study an arbitrary sector with momentum ${\bf K}$, 
satisfying the following relation:
\begin{equation}\label{eq:bloch_property}
    \text{Log}[\Psi^K_{\theta}(T_{\bf R}\sigma)] = \text{Log}[\Psi^K_{\theta}(\sigma)] + i {\bf K} \cdot {\bf R} \ ,
\end{equation}
where $T_{\bf R}$ represents a translation by the Bravais lattice vector ${\bf R}$, whose components are integer multiples of the primitive vectors ${\bf a}_1$
and ${\bf a}_2$. The momenta ${\bf K}$ are quantized in units of $2\pi/L$, i.e, $K_x=2\pi n/L$, with $n$ ranging from $0$ to $L/2-1$ and similarly for $K_y$. 
To define the ViT state with a specific momentum ${\bf K}$, we modified the amplitude as follows:
\begin{equation}\label{eq:psi_k}
    \text{Log}[\Psi^K_{\theta}(\sigma)]= \text{Log}[\Psi_{\theta}(\sigma)] + i \Theta_K(\sigma) \ ,
\end{equation}
where the function $\Theta_K(\sigma)$ adjusts the phase to match the target momentum sector, ensuring that Eq.~\eqref{eq:bloch_property} is satisfied.
This phase shift can be computed from the output vectors $(\boldsymbol{y}_1, \dots, \boldsymbol{y}_n)$ as follows:
\begin{equation}\label{eq:theta_k}
    \Theta_K(\sigma) = \Im\left\{\text{Log}\left(\sum_{j=1}^{n} e^{i {\bf K} \cdot {\bf R}_j}{y}_{j,1}\right)\right\} \ .
\end{equation}
where ${\bf R}_j$ indicates the (Bravais) vector that identifies the $j$-th patch of the Shastry-Sutherland model and $y_{j,1}$ is the first component of the 
vector $\boldsymbol{y}_j$, chosen by convention. 

\begin{figure}[t]
\center
\includegraphics[width=\columnwidth]{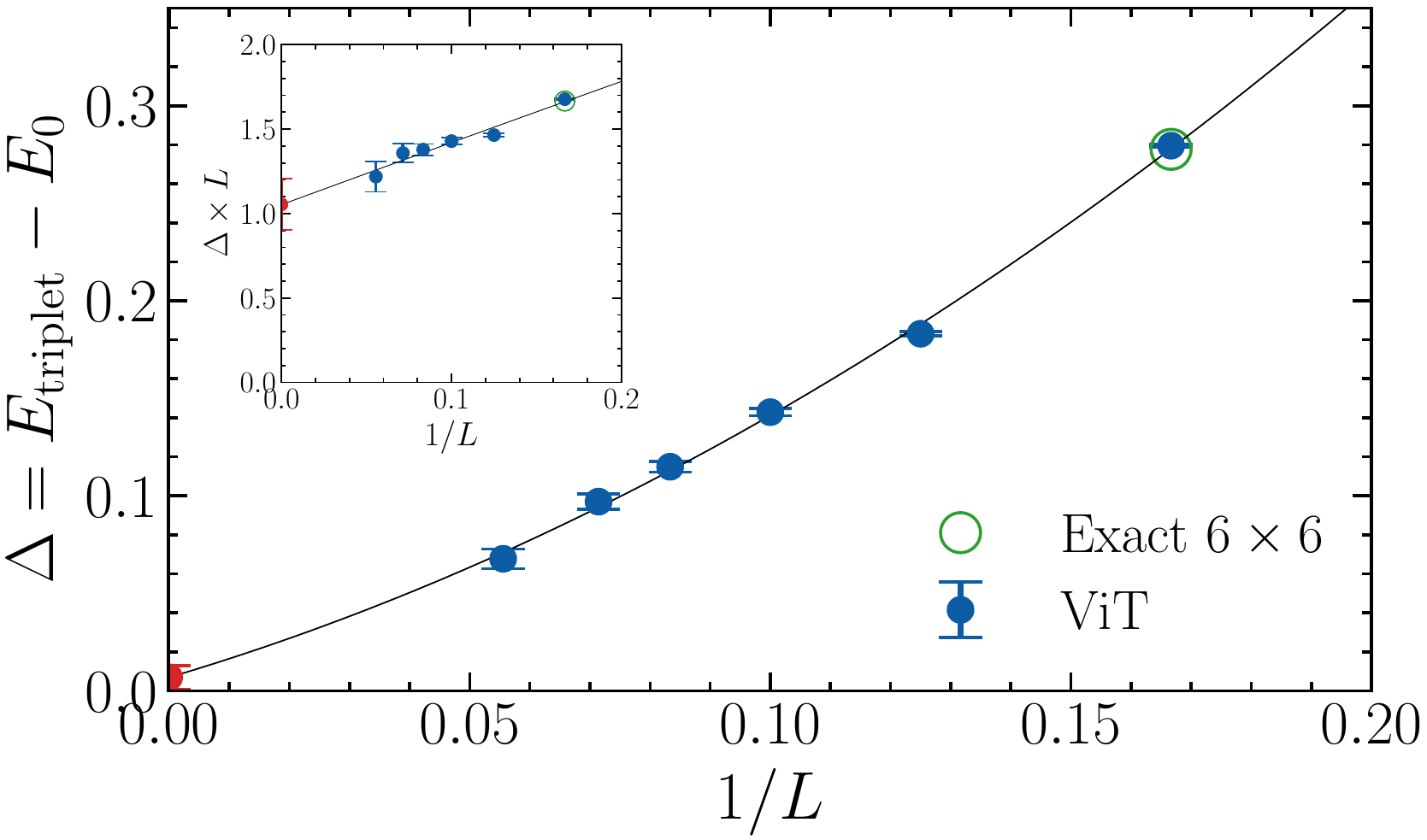}
\caption{\label{fig:gap} 
Energy gap $\Delta$ obtained by the ViT wave function between the ground state and the lowest-energy triplet state, as a function of inverse linear length $1/L$,
from $L=6$ to $L=18$ at $J/J^\prime=0.8$. The exact gap for the $6 \times 6$ lattice is also reported for comparison (green empty circle). Inset: Rescaled gap 
$\Delta \times L$ as a function of $1/L$ for the same size considered in the main panel.}
\end{figure}

In practice, we fix one of the possible momenta allowed in the $L \times L$ lattice and perform the Monte Carlo sampling in the sector with $S^z=1$. Even though
the latter condition does not imply that the variational state is a triplet, we verified that the expectation value of the total spin $S^2$ is very close to $2$.
Therefore, this {\it Ansatz} gives an accurate approximation of a triplet state. The left panel of Fig.~\ref{fig:excited} shows the optimization curves of the 
variational energy for the three independent momenta of the $6 \times 6$ cluster, namely ${\bf K}a = (0,0)$, $(0, 2\pi/3)$, and $(2\pi/3, 2\pi/3)$ (where $a=2$
is the length of the primitive vectors, to have $-\pi < K_xa \le \pi$, and similarly for $K_y$). In all the cases, the ViT wave function achieves a relative error 
of approximately $\Delta\varepsilon \approx 10^{-3}$ within $10^4$ optimization steps.  The resulting variational triplet energy gaps have high accuracy when 
compared with the exact ones (refer to the right panel of Fig.~\ref{fig:excited}). Specifically, we obtain that the lowest-energy excitation on the $6 \times 6$ 
lattice lies at the $\Gamma$ point. In the right panel of Fig.~\ref{fig:excited}, we also report the triplet gaps for a closed path in momentum space on a 
$12 \times 12$ cluster, confirming that the zero-momentum excitation remains the lowest-energy one.

Then, we proceed to analyze the size scaling of the lowest-energy triplet at $J/J^\prime=0.8$. The energy gap ${\Delta = E_{\text{triplet}} - E_0}$ is reported in
Fig.~\ref{fig:gap} for different values of the cluster size, i.e., with $L$ ranging from $6$ to $18$. The extrapolation performed using a quadratic fit of the form 
$\Delta = a + b/L + c/L^2$ yields, with a small fitting error, a vanishing gap in the thermodynamic limit, i.e., $\Delta=0.00(7)$. In addition, in the inset of 
Fig.~\ref{fig:gap}, we show that the rescaled gap $\Delta \times L$ approaches a finite value in the thermodynamic limit, thus confirming the gapless nature of
the intermediate spin-liquid phase.

\subsection{Hidden representations}\label{subsec:hidden_rep}

We discussed in Section~\ref{sec:wavefunction} that the motivation for our ViT wave function is to leverage the power of representation learning. Instead of
using a neural network simply as a universal approximator to map spin configurations to wave function amplitudes, we train the network to map spin states into 
a feature space. These features are then used as an input to a shallow neural network to predict the amplitude corresponding to the input configuration [see 
Eq.~\eqref{eq:rbm}]. This change in perspective has important practical ramifications: it dispenses with the need of having complex-valued parameters in the 
feature extractor $V(\boldsymbol{\sigma}; \phi)$, making the training of deep networks much easier. In this section, we verify that the network indeed learns 
a set of non-trivial features in the course of training, and we find that some of these features are even interpretable in a simple limiting case.

\begin{figure}[t]
\includegraphics[width=0.9\linewidth]{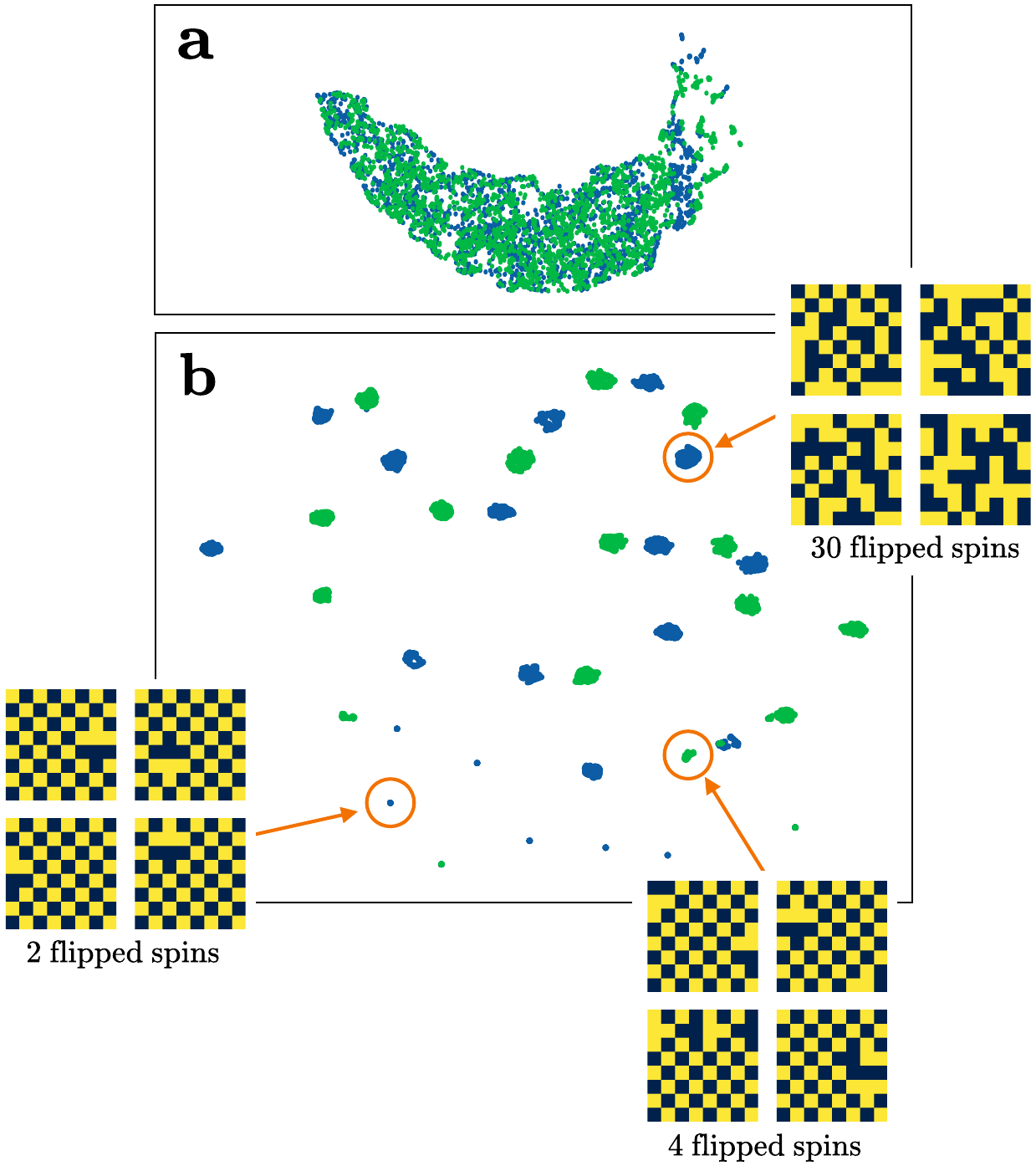}
\caption{\label{fig:umap}
Dimensional reduction (using the UMAP algorithm) of the hidden representations for a set of configurations obtained from a ViT in the limit of $J^\prime=0$, leading to the Heisenberg model. 
Points are colored according to the exact signs given by the Marshall sign rule~\cite{marshall1955}.
a) Before optimizing, when the parameters of the neural network are random, the hidden representations are all concentrated in a single cluster. b) After the optimization of the variational energy the hidden representations are organized in clusters with the same number of flipped spins with respect to the Néel state. All calculations are performed on an $8 \times 8$ cluster.}
\end{figure}

\begin{figure}[t]
\includegraphics[width=0.8\columnwidth]{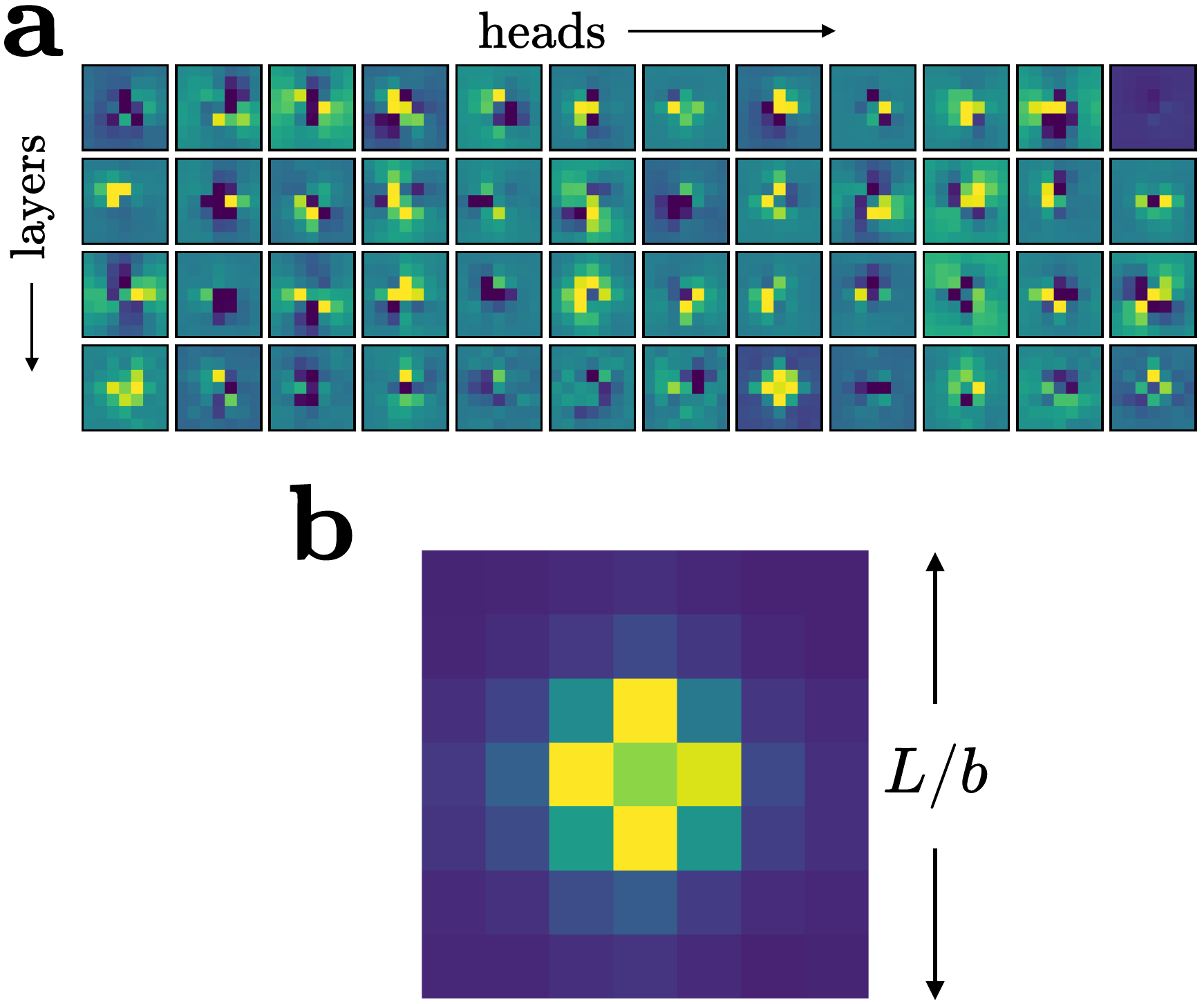}
\caption{\label{fig:attention_maps}
a) Attentions maps of a ViT with 4 layers and 12 heads per layer, optimized in the point $J/J^\prime=0.8$ for $L=14$. b) Mean of the absolute values of all the 
attention maps. The attention maps have size $L/b$, where $L$ and $b$ are the linear dimensions of the lattice and of the patches, respectively.}
\end{figure}

To this end, we examine the limit $J^\prime=0$, in which the system reduces to the unfrustrated Heisenberg model; here, the ground state properties are characterized 
by the presence of antiferromagnetic order. In addition, the exact sign structure of the ground state follows the Marshall sign rule~\cite{marshall1955}. For a given 
set of configurations $\{ \boldsymbol{\sigma}_i \}$ (sampled along the Monte Carlo procedure), we compute the corresponding hidden vectors $\{\mathbf{z}_i\}$ of size 
$d \gg 1$, which can be visualized in two dimensions after a dimensional
reduction with the standard Uniform Manifold Approximation and Projection 
method (UMAP)~\cite{mcinnes2020}.

In Fig.~\ref{fig:umap}, we color each feature or representation $\mathbf{z}_i$ according to the exact sign of the amplitude corresponding to the spin configuration 
$\boldsymbol{\sigma}_i$. Before the Transformer is trained, i.e., with random network weights, there is no discernible structure in the representations (see 
Fig.~\ref{fig:umap}a). After training, by means of minimizing the variational energy, we see instead that features have a highly non-trivial structure: they are 
grouped into different clusters of the representation space (see Fig.~\ref{fig:umap}b). This is a direct result of the training.

Remarkably, these clusters have a physical interpretation in the unfrustrated case, where the spin configurations in a given cluster have the same number of flipped 
spins with respect to the N\'eel state and, therefore, the same sign (according to the Marshall rule), and similar modulus. The crucial point is that, by using a
single fully-connected layer, the prediction of the correct amplitudes is much easier when acting on these representations rather than using the original spin
configurations, as can be seen from the vastly superior energies obtained by the ViT state compared to a simple RBM~\cite{rende2023,rende2024finetuning}.

A similar clustering structure in the trained ViT also appears for general values of $J/J^\prime$, though the precise interpretation of the clusters is less 
straightforward (see Appendix~\ref{app:hidden_rep}), and is left for future works. This reflects the situation in machine learning, where it is generally difficult 
to extract human-interpretable structure directly from the representations. In fact, the whole point of letting neural networks learn features directly from data 
is that they can discover subtle patterns in the data that are hard for humans to extract or even describe, but nevertheless turn out to be useful for the task at 
hand.

\subsection{Attention maps}\label{subsec:att_maps}

Another approach to understand how the ViT wave function processes the input spin configurations is to examine the attention weights $\alpha_{i-j}$ of a trained 
Transformer for the different heads and layers at $J/J^\prime=0.8$ and $L=14$, which we show in Fig.~\ref{fig:attention_maps}a. A key feature of the self-attention 
mechanism is to connect all the input patches even in a single layer. We highlight that the network makes use of this capability even in the first layer, since 
some heads attend to all the patches. This is not possible when working with architectures that use only local filters (e.g., convolutional ones). To capture the 
overall behavior of the attention weights, we compute the mean of the absolute values of the weights across all heads and layers. The results, displayed in 
Fig.~\ref{fig:attention_maps}b, reveal a systematic trend: the mean interaction between patches (representing groups of spins in this context) exhibits a regular 
decay as the relative distance between patches increases. Interestingly, this mean attention map encodes also the rotational symmetry of the model, which is not 
imposed in the weights, whereas individual attention maps do not exhibit this feature (see Fig.~\ref{fig:attention_maps}a). These observations are fully consistent 
with the findings of Ref.~\cite{queries_keys}, in which analytical results establish a direct link between attention weights and correlations among spins.

\section{Conclusions}

Our results demonstrate that NQSs represent an extremely useful tool to investigate the ground-state properties of frustrated quantum magnets. Here, we focused 
the attention on the Shastry-Sutherland model, for which the existence of a spin-liquid phase between the plaquette and antiferromagnetic ones has been recently 
suggested~\cite{yang2022,wang2022}. The difficulty of the problem resides in the smallness of this region, thus requiring extremely accurate calculations and 
large system sizes. The present definition of the ViT wave function (that combines a real-valued attention mechanism and a final complex-valued fully-connected 
layer) allows us to detect the existence of a finite region $0.78 \lesssim J/J^\prime \lesssim 0.82$ in which both magnetic and plaquette orders vanish in the 
thermodynamic limit, thus supporting the presence of the intermediate spin-liquid phase~\cite{yang2022}. Most importantly, by a direct evaluation of the energy 
gap between the ground state and the lowest-energy triplet excitation, we provided evidence for a {\it gapless} spin-liquid phase. Remarkably, the characterization 
of the spin liquid phase within the Shastry-Sutherland model has not been explored in previous studies. Our results are particularly important because they show 
that the magnetically ordered N\'eel phase is melted into a gapless spin liquid, similarly to what happens in the $J_1$-$J_2$ Heisenberg model on the square 
lattice~\cite{hu2013}. This suggests that this kind of (continuous) transition is rather generic and may represent the habit, and not the exception, for the 
melting of the N\'eel order due to magnetic frustration. In addition, our calculations clearly demonstrate that the ViT {\it Ansatz} rises among the universe of 
variational wave functions as a possible way to eventually solve important quantum many-body problems. One key feature is the ability of this approach to create 
a mapping of the physical configurations in a {\it real} feature space, where it is then easy to predict amplitudes, even with a single fully-connected layer. 
Looking at NQSs as feature extractors is another original contribution of this work, in contrast with the common interpretation of just universal approximators 
of functions, which usually leads to taking all the parameters complex-valued.

Future directions are two-fold. From the physical point of view, it is tantalizing to apply this approach to other many-body problems, including fermionic systems,
which pose the challenge of grasping the correct antisymmetry of the wave function. In these cases, at present NQS do not achieve comparable accuracies as observed 
in spin models, underscoring a rich area for improvement and exploration. From the machine-learning part, the matter for future research would be an examination of 
the attention maps learned by the ViT, checking whether they could be used to directly infer physical properties of the ground state, without the need to compute order 
parameters. Moreover, it could be interesting to study in detail the representations (clusters) built by the Transformer, in particular how they change across the 
different layers and in the different phases, in such a way as to understand phase transitions by looking only at hidden representations. Along this line of research, 
it would be valuable to investigate whether the clusters identified when working with the $z$-axis basis can be utilized to detect orders of off-diagonal operators.

\begin{acknowledgments}
We thank S. Sachdev, A. Sandvik, M. Imada, A. Chernyshev, A. Laio and Y. Iqbal for useful discussions. We also acknowledge L. Wang for providing us the DMRG energies 
from Ref.~\cite{yang2022}. We acknowledge the CINECA award under the ISCRA initiative, for the availability of high-performance computing resources and support.
\end{acknowledgments}

\appendix

\section{Lattice and symmetries}\label{app:symmetries}

\begin{figure}[t]
\includegraphics[width=0.6\columnwidth]{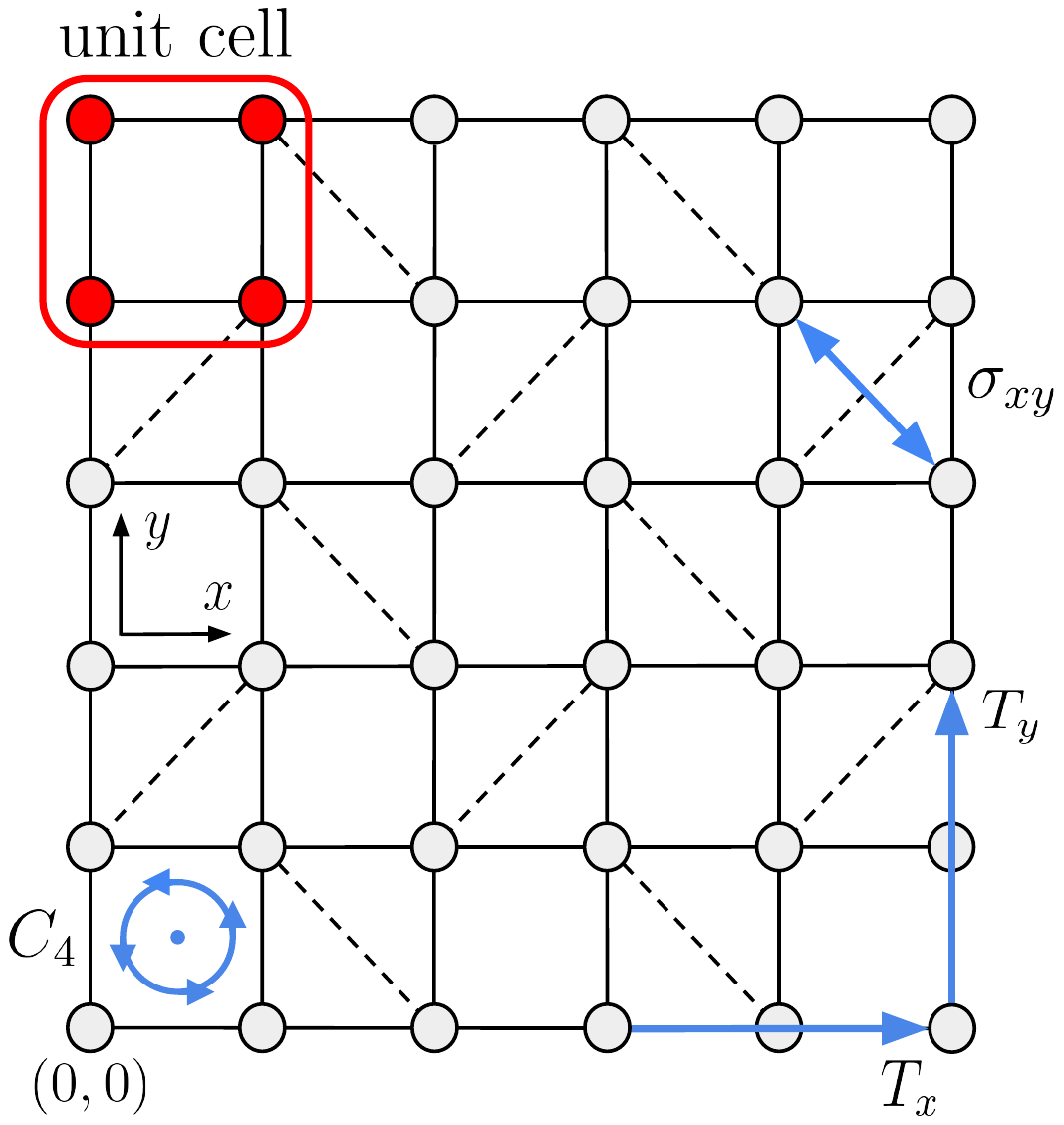}
\caption{\label{fig:lattice_sym} 
The (nearest-neighbor) coupling $J$ is denoted by solid lines and (next-nearest-neighbor one) $J^\prime$ by dashed lines. The standard unit cell contains $4$ sites, 
implying translations $T_x$ and $T_y$ (along $x$ and $y$ axis) by $2$ lattice points. The point-group symmetries, $C_4$ rotations and $\sigma_{xy}$ reflection, are 
also shown.}
\end{figure}

The Shastry-Sutherland lattice is shown in Fig.~\ref{fig:lattice_sym}, where each site is labeled by the Cartesian coordinate ${\bf r}=a/2(x,y)$, where $x$ and 
$y$ are integers and $a=2$ is the length of the primitive lattice vectors ${\bf a}_1=(a,0)$ and ${\bf a}_2=(0,a)$. The lattice is invariant under translations 
$T_{{\bf a}_1}: (x,y) \rightarrow (x+a,y)$ and $T_{{\bf a}_2}: (x,y) \rightarrow (x, y+a)$. This symmetry can be easily encoded in the Transformer architecture 
by taking as input patches the four spins in an empty plaquette (i.e., plaquettes with no $J^\prime$ bonds), which constitute the unit cell and then choosing 
the translationally invariant attention weights, namely $\alpha_{i,j}=\alpha_{i-j}$. In addition, the lattice is invariant under the rotation with respect to 
the center of the empty plaquette at the origin of the lattice $R_{\pi/2}: (x,y) \rightarrow (-y+1,x)$ and the diagonal reflection 
$\sigma_{xy}: (x,y) \rightarrow  (y+1,x-1)$. For the ground state, which lies in the ${\bf K}a=(0,0)$ sector, all these symmetries can be enforced by a projector 
operator, leading to a total-symmetric state~\cite{nomuraimada2021,nomura2021,moritzschmitt2023}:
\begin{equation}\label{eq:symm_psi}
    \tilde{\Psi}_{\theta}(\sigma) = \sum_{{\rho}, {\cal R}} \Psi_{\theta}({\rho} {\cal R} \sigma),
\end{equation}
where ${\rho} \in \{\mathbb{I}, \sigma_{xy}\}$ and ${\cal R} \in \{\mathbb{I}, R_{\pi/2}, R^2_{\pi/2}, R^3_{\pi/2}\}$. Notice that the sum is over a fixed number 
of terms and does {\it not} scale with the size of the system. In general, this procedure gets an improvement in the accuracy of the variational state, which is 
difficult to obtain by just increasing the number of variational parameters. The numerical simulations shown in this work are performed with the symmetrized state 
in Eq.~\eqref{eq:symm_psi}.

\section{Optimization of the variational parameters}\label{app:optimization}

The standard formulation of the Stochastic Reconfiguration~\cite{sorella2005,becca2017} requires inverting a square matrix whose size is equal to the number 
of variational parameters. The computational cost of this matrix inversion is prohibitive when increasing the number of parameters and limits this approach to a 
relatively small number of parameters compared to modern deep learning models. However, two recent papers~\cite{chen2023,rende2023} proposed variations of the 
original algorithm that can deal with variational states with millions of parameters $P$, working in the regime where $P$ exceeds the number of samples $M$ used 
for the stochastic estimations. These approaches lead to the following updates:
\begin{equation}\label{eq:sr2}
    \bm{\delta \theta} = \tau X(X^TX + \lambda \mathbb{I}_{2M})^{-1}\bm{f} \ ,
\end{equation}
where $\tau$ is the learning rate and $\lambda$ is the regularization parameter. The matrix $X$ has shape $P \times 2M$ and it is obtained as the concatenation of 
the real and imaginary parts of the centered rescaled Jacobian ${Y_{\alpha,i} = (O_{\alpha i} - \bar{O}_{\alpha})/\sqrt{M}}$, where 
${O_{\alpha,i} = {\partial}\text{Log}[\Psi_{{\theta}}(\sigma_i)]/{\partial \theta_{\alpha}}}$ are the logarithmic derivatives. The vector 
$\boldsymbol{f} \in \mathbb{R}^{2M}$ is given by ${\boldsymbol{f}= \text{Concat}[\Re(\bm{\varepsilon}),-\Im(\bm{\varepsilon})]}$, having introduced the centered 
rescaled local energy $\varepsilon_{i}=-2[E_{L,i}- \bar{E}_L]^*/\sqrt{M}$, with ${E_{L,i} = \braket{\sigma_i|\hat{H}|\Psi_{\theta}}/\braket{\sigma_i|\Psi_{\theta}}}$. 
The expressions $\bar{E}_L$ and $\bar{O}_\alpha$ are used to denote sample means. A detailed derivation of the Eq.~\eqref{eq:sr2} can be found in Ref.~\cite{rende2023}.

\begin{figure}[t]
\includegraphics[width=\columnwidth]{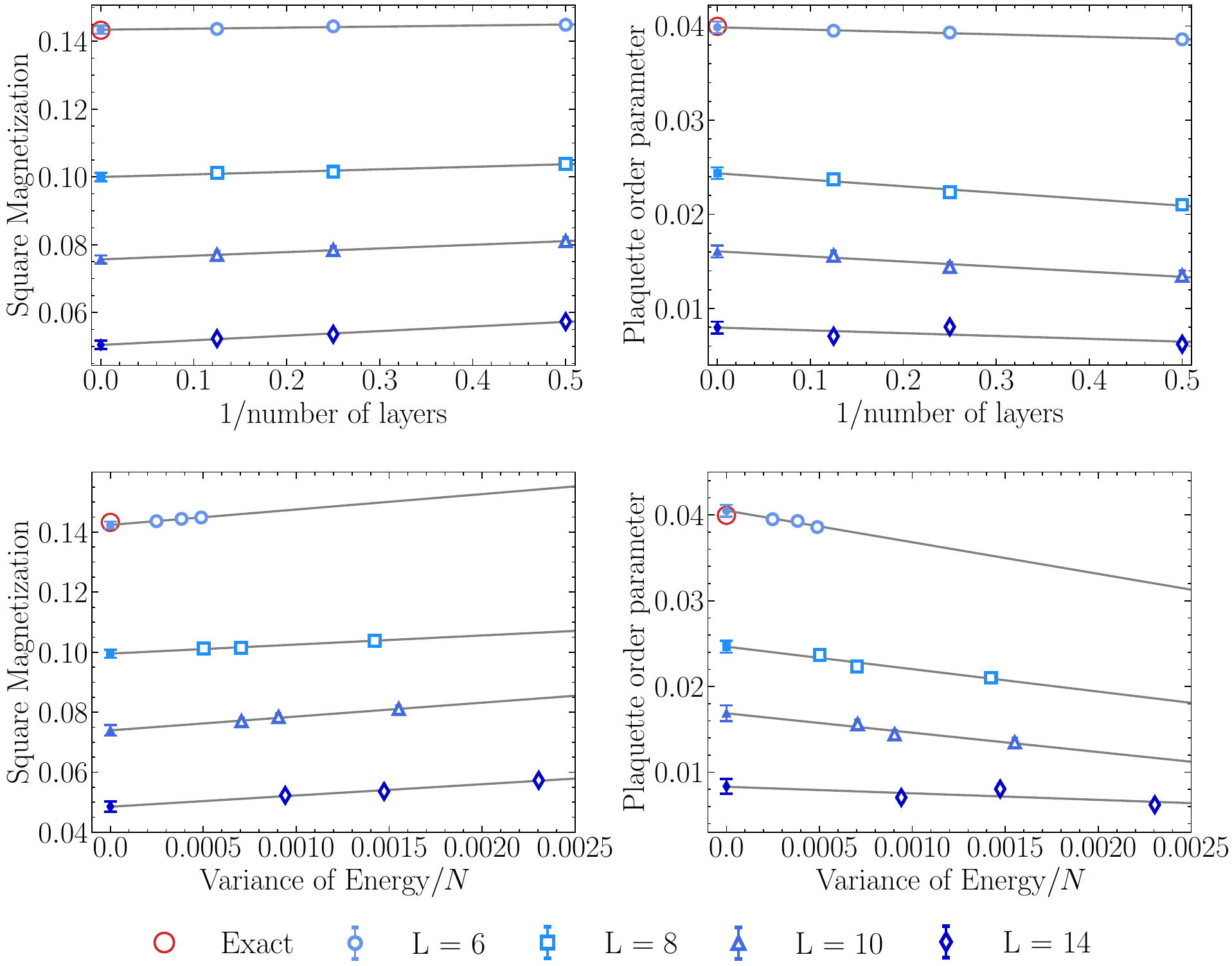}
\caption{\label{fig:extrapolations}
Upper panels: the values of the square magnetization $m^2(L)$ (left panel) and of the plaquette order parameter $m_p(L)$ (right panel) as a function of the number of 
layers $n_l=2,4,8$ and for different lattice sizes from $L=6$ to $L=14$ at $J/J^\prime=0.8$. Lower panels: the same quantities plotted as function of the variance
of the energy (divided by $N$). Extrapolated results for the limit of an infinite number of layers and in the limit of zero variance are displayed as filled symbols, 
with exact values for $L=6$ shown for comparison (red circles). Error bars on the extrapolated values are determined using resampling techniques under the assumption 
of Gaussian noise.}
\end{figure}

This formulation of the Stochastic Reconfiguration is implemented in NetKet~\cite{netket3:2022}, under the name of 
\href{https://netket.readthedocs.io/en/latest/api/_generated/experimental/driver/netket.experimental.driver.VMC_SRt.html#netket.experimental.driver.VMC_SRt}{\texttt{VMC\_SRt}}.

\section{Extrapolations details}\label{app:extrapol}

\begin{figure}[t]
\includegraphics[width=\columnwidth]{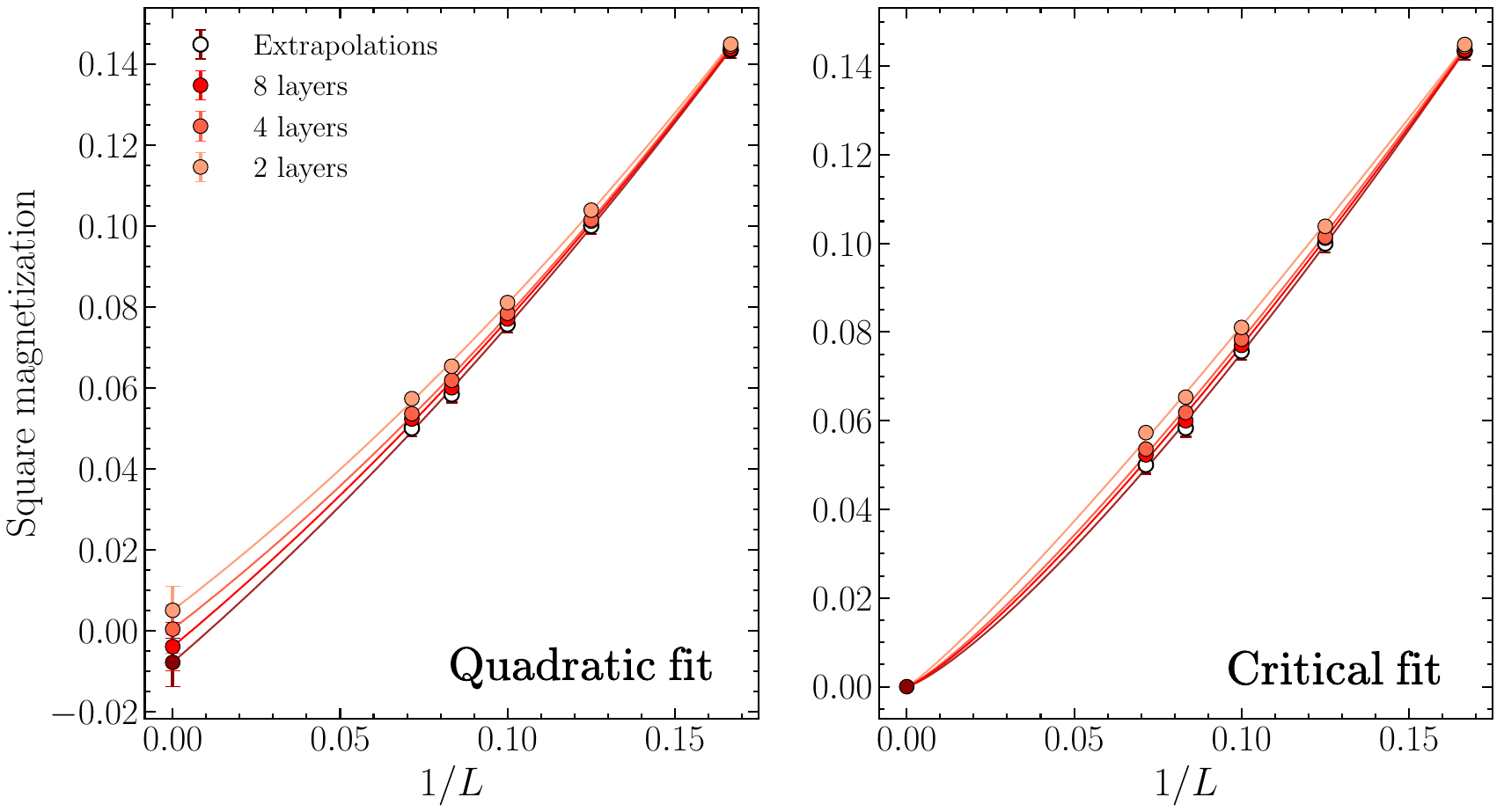}
\caption{\label{fig:extr_appendix} 
Size scaling of the square magnetization $m^2(L)$ as a function of $1/L$ from $L=6$ up to $L=14$ at $J^\prime/J=0.8$. The numerical data encompasses varying numbers 
of layers, specifically from $n_l=2$ to $n_l=8$, along with the extrapolated values for an infinite number of layers  (see Fig.~\ref{fig:extrapolations}). The curves 
for the extrapolations in the thermodynamic limit are performed using as fitting curve a second-order polynomial in $1/L$ [see Eq.~\eqref{eq:quadratic_fitting}] (left 
panel) and the critical form of Eq.~\eqref{eq:critical_fitting} (right panel). The error bars of the extrapolated values in the thermodynamic limit are obtained with 
resampling techniques with gaussian noise.}
\end{figure}

Here, we provide further details on the extrapolation procedures used to obtain the final values of the order parameters presented in Fig.~\ref{fig:phase_transition} 
of the main text. In the upper panels of Fig.~\ref{fig:extrapolations}, we show the order parameters, namely the square magnetization $m^2(L)$ and the plaquette 
order parameter $m_p(L)$, plotted as a function of the number of layers $n_l$, extrapolating their values for a network with $n_l \to \infty$. For $L=6$, these 
numerical extrapolations show excellent agreement with exact diagonalization results. Furthermore, the extrapolated values exhibit minimal deviation from those 
obtained with $n_l=8$ layers, underscoring the consistency of the results. To further assess the robustness of the calculations, in the lower panels of 
Fig.~\ref{fig:extrapolations} the extrapolations are also performed as a function of the variance of the energy. Notably, this alternative method, which does not 
depend on the specific structure of the variational state, yields results consistent with the layer-based extrapolations. Finally, in Fig.~\ref{fig:extr_appendix}, 
we cross-validate the results by utilizing both a second-order polynomial in $1/L$ [see Eq.~\eqref{eq:quadratic_fitting}] and the critical form described in 
Eq.~\eqref{eq:critical_fitting} for the extrapolations of the square magnetization in the thermodynamic limit (as a function of $1/L$). We repeat the extrapolations 
considering each value of the number of layers $n_l=2$, $4$, and $8$. The resulting fitting curves exhibit remarkably similar behaviors, further confirming the 
consistency and reliability of the extrapolated results.

Table~\ref{table:energies} reports the ground-state variational energies per site (in units of $J^\prime$) for system sizes ranging from $L=6$ to $L=14$, computed 
using a Transformer architecture with $n_l$ layers, where $n_l=2$, $4$, and $8$. The last column shows the extrapolated energies for an infinite number of layers, 
obtained through variance extrapolation technique~\cite{hu2013}. Additionally, the ground-state energies for $ n_l = 8 $ are provided for the largest system sizes 
considered in this work, $L=16$ and $L=18$.

\begin{table}[b]
\begin{tabular}{lllll}
\hline \multicolumn{1}{c}{L}
 & \multicolumn{1}{c}{2 layers} & \multicolumn{1}{c}{4 layers} & \multicolumn{1}{c}{8 layers} & \multicolumn{1}{c}{Extrap.} \\ \hline
\multicolumn{1}{c}{6}   & -0.4516642 & -0.4516991 & -0.4517072 & -0.451750  \\
\multicolumn{1}{c}{8}   & -0.449641  & -0.449802  & -0.449829  & -0.44995   \\
\multicolumn{1}{c}{10}  & -0.449062  & -0.449221  & -0.449329  & -0.44947   \\
\multicolumn{1}{c}{12}  & -0.448861  & -0.449013  & -0.449078  & -0.44931   \\
\multicolumn{1}{c}{14}  & -0.448551  & -0.448812  & -0.448929  & -0.44920   \\
\multicolumn{1}{c}{16}  &           &           & -0.448881  &            \\
\multicolumn{1}{c}{18}  &           &           & -0.448859  &            \\
\hline \hline
\end{tabular}
\caption{\label{table:energies} Ground-state variational energy (in unit of $J^\prime$) for different number of layers $n_l$ at  $J/J^\prime=0.8$. The extrapolated 
values obtained by variance extrapolation~\cite{nomura2021} for an infinite number of layers are also reported. The Monte Carlo error due to finite sampling effects 
is on the last digit. In the case of a $6\times 6$ lattice, the ground-state energy per site from exact diagonalization is $E=-0.4517531$.}
\end{table}

\section{Study of large lattice sizes}\label{app:large_lattices}

The ViT architecture outlined in Section~\ref{sec:wavefunction}, due to the fully connected structure of the attention mechanism [see Eq.~\eqref{eq:fact_attention}], 
exhibits a computational complexity that scales quadratically with the input sequence length. This quadratic scaling leads to substantial computational costs when 
increasing the size of the system. Here, we propose two alternative approaches to mitigate this computational challenge.

\begin{figure}[t]
\includegraphics[width=\columnwidth]{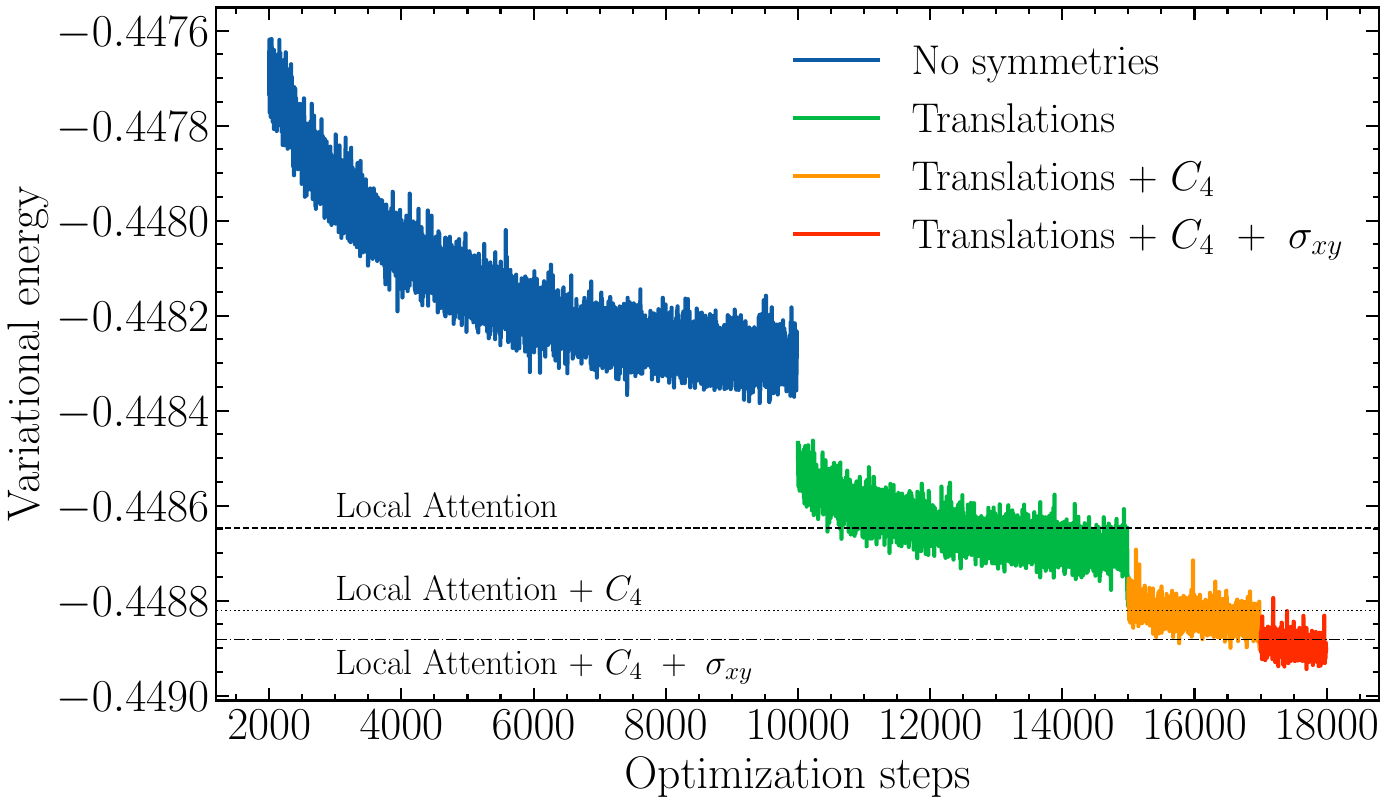}
\caption{\label{fig:loss_b4}
Variational energy as a function of optimization steps for a ViT optimized to approximate the ground state of a $16 \times 16$ lattice at coupling ratio 
$J/J^\prime=0.8$. The hyperparameters of the network are set as follows: attention heads $h = 12$, embedding dimension $d = 72$, number of layers $n_l = 12$, 
and a patch size of $b = 4$. For enhanced visual clarity, the initial $2 \times 10^3$ steps have been omitted from the plot. Horizontal dashed lines indicate 
the final energies obtained using a local attention mechanism with a patch size of $b = 2$ and a window size restricted to $5 \times 5$ patches. For this local 
attention setup, the network’s hyperparameters are $h = 12$, $d = 72$, and $n_l = 8$. In both setups, the optimization protocol employs a sample size of 
$M = 6\times 10^3$, a learning rate of $\tau = 0.03$, and a diagonal shift regularization parameter of $\lambda = 10^{-4}$ (see Appendix~\ref{app:optimization} 
for details).}
\end{figure}

The first approach utilizes {\it local} attention mechanisms, which compute the updated representation of a patch based only on its neighboring patches. This method 
is employed in our simulations of the $16 \times 16$ and $18 \times 18$ lattice systems. Specifically, for these larger sizes, we utilize a network with hyperparameters 
$h=12$, $d=72$, $n_l=8$, and a local attention window size of $5 \times 5$ patches. In this case, we do not perform the extrapolations of the order parameters in the 
number of layers: the robustness of the values obtained for a number of layers $n_l=8$ is justified by the analysis presented in Appendix~\ref{app:extrapol}. 
We emphasize that the use of local attention preserves the translational invariance of the variational state.

The second method to reduce the computational complexity with the input sequence length is to increase the patch size $b$. This method has the advantage of preserving 
global connections of the patches in each layer but has the drawback of breaking translational symmetries, which must subsequently be restored, as with other symmetries 
(see Appendix~\ref{app:symmetries}).

In Fig.~\ref{fig:loss_b4}, we present the optimization curves of the variational energy for a $16 \times 16$ lattice, employing a neural network architecture with 
hyperparameters $h=12$, $d=72$, and $n_l=12$, and utilizing a patch size of $b=4$. The optimization proceeds in stages: an initial phase of $10^4$ steps is performed 
without enforcing symmetries, during which the ViT state retains only translational symmetry among patches. This first stage is followed by the restoration of 
translational symmetry over $4\times 10^3$ steps. Subsequently, $C_4$ rotational symmetry and reflection symmetry are imposed, optimized over $2\times 10^3$ and 
$10^3$ steps, respectively. For comparison, horizontal dashed lines represent the final energies obtained using a local attention approach with a patch size of 
$b=2$ and hyperparameters $h=12$, $d=72$, $n_l=8$. The close agreement between the final energies across both approaches highlights the robustness and consistency 
of the variational results across the two different setups.\\

\begin{figure}[t]
\includegraphics[width=\columnwidth]{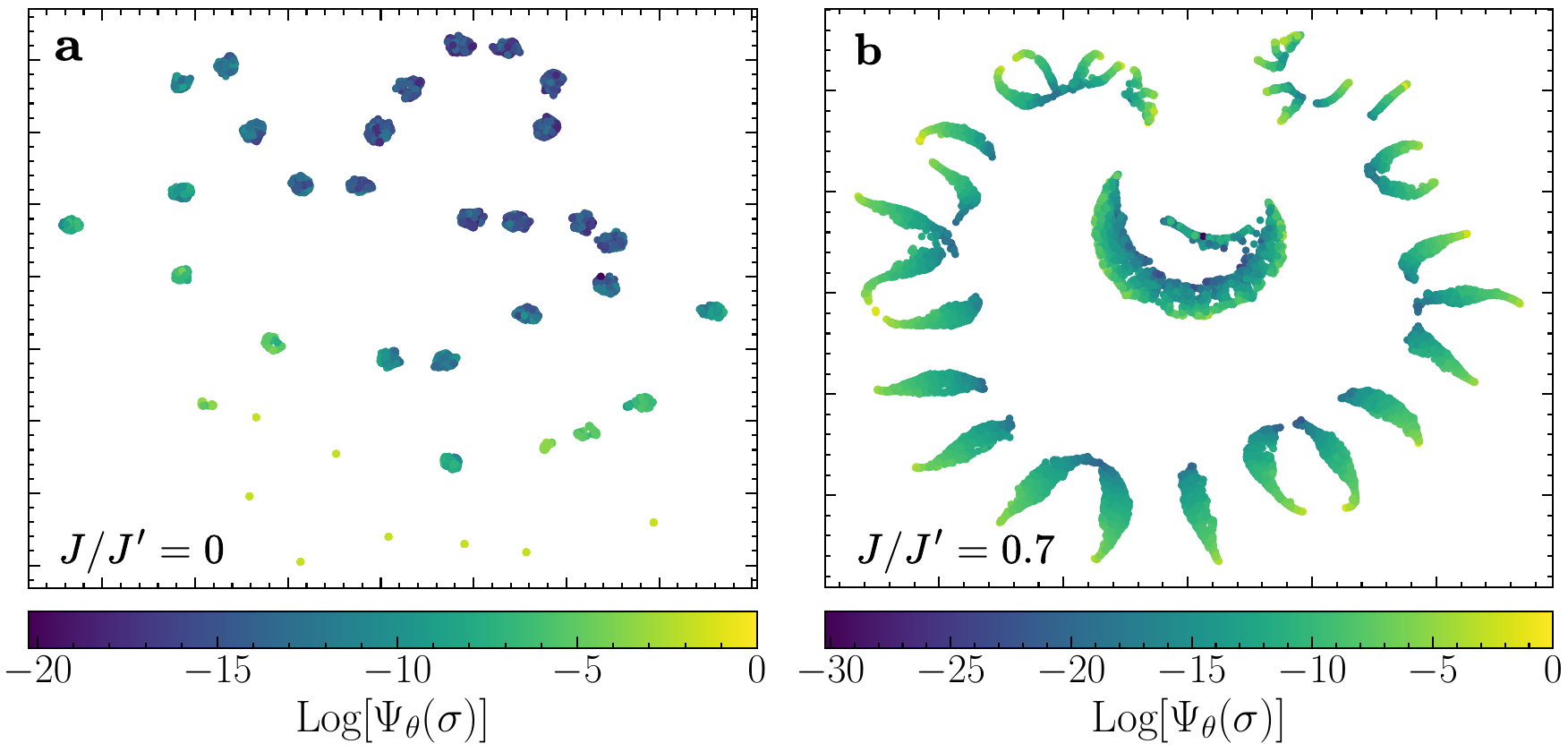}
\caption{\label{fig:umap_frustr} 
UMAP projections of the hidden representations, at the end of the variational energy minimization, for the unfrustrated [panel (a)] and frustrated case with 
$J/J^{\prime}=0.7$ [panel (b)] on a $8 \times 8$ cluster. Points are colored according to the log-amplitude of the corresponding spin configurations, normalized 
with respect to the maximum amplitude within the sample.}
\end{figure}

\section{Hidden Representations in Frustrated scenario}\label{app:hidden_rep}

In subsection~\ref{subsec:hidden_rep}, we discussed the role of the hidden representations in determining the effectiveness of our variational {\it Ansatz} [see 
Eq.~\eqref{eq:composition}]. In Fig.~\ref{fig:umap_frustr}, we compare UMAP projections for the unfrustrated case and the frustrated case with $J/J^{\prime}=0.7$. In these plots, the hidden representations $\boldsymbol{z}_i$ are visualized with colors assigned according to the predicted logarithmic amplitude of the associated configurations $\boldsymbol{\sigma}_i$, which is normalized relative to the maximum amplitude within the sample. Importantly, we observe that even for a generic frustration ratio  $J/J^{\prime}$ , a clustering structure consistently emerges in the feature space at the end of the energy minimization. This outcome validates the representation learning framework that inspired the design of our \textit{Ansatz}. 
Despite starting from the same random initialization of the variational parameters, resulting in all configurations being concentrated within a single cluster (see Fig.~\ref{fig:umap}a), the feature spaces obtained after the energy minimization exhibit significant differences. For the unfrustrated case, there is a global gradient in the amplitudes. Points in the bottom clusters correspond to configurations with a small number of spin flips relative to the N\'eel state, resulting in larger amplitudes (see Fig.~\ref{fig:umap}). Moving toward the top clusters, the number of spin flips increases, leading to a decay in amplitude. In contrast, for the frustrated case, the arrangement of configurations in the feature space is significantly different. Configurations mapped within the same cluster display very different amplitudes, with each cluster exhibiting its own internal gradient of amplitudes. This complication makes interpreting the clustering structure more challenging compared to the unfrustrated scenario.
\bibliography{refs}

\end{document}